\documentclass[preprint,12pt,sort&compress]{elsarticle}

\usepackage{amssymb}
\usepackage{amsmath}
\usepackage{color}
\usepackage{float} 

\journal{Physics of the Dark Universe}

\begin{document}
	
	\begin{frontmatter}
		
		
		
		\title{Impact of Perfect Fluid Dark Matter on the Appearance of Rotating Black Hole}
		
		
		\author[1]{Yu-Xiang Huang} 
		\ead{yxhuangphys@126.com}
		\author[2]{Sen Guo\corref{cor1}}
		\ead{sguophys@126.com}
		\author[1]{En-Wei Liang\corref{cor1}}
		\ead{lew@gxu.edu.cn}
		\author[3]{Kai Lin}
		\ead{kailin@if.usp.br}

		\cortext[cor1]{Corresponding author}
		
		\affiliation[1]{
			organization={Guangxi Key Laboratory for Relativistic Astrophysics, School of Physical Science and Technology, Guangxi University},
			city={Nanning},
			postcode={530004},
			country={People's Republic of China}
		}

		\affiliation[2]{
			organization={College of Physics and Electronic Engineering, Chongqing Normal University},
			city={Chongqing},
			postcode={401331},
			country={People's Republic of China}
		}

		\affiliation[3]{
			organization={Universidade Federal de Campina Grande},
			city={Campina Grande},
			postcode={58429-900},
			state={PB},
			country={Brasil}
		}
		
		\begin{abstract}
			Understanding how dark matter affects the immediate environment of black holes (BHs) is crucial for interpreting horizon-scale observations. We study rotating BHs surrounded by perfect fluid dark matter (PFDM), exploring their observable features through both analytical and numerical approaches. Using the existence criterion of the innermost stable circular orbit (ISCO), we first derive joint constraints on the PFDM intensity parameter~$k$ and the spin parameter~$a$. Within the resulting physically allowed parameter regime, we perform high-resolution, general-relativistic ray-tracing simulations of thin accretion disks at 87~GHz and 230~GHz, capturing the detailed brightness morphology and photon ring structure shaped by PFDM. By incorporating angular diameter measurements of M87$^{*}$ and Sgr~A$^{*}$ from the Event Horizon Telescope (EHT), we further narrow down the viable parameter space and directly compare synthetic images with EHT observations of M87$^{*}$. We find that the inclusion of PFDM improves the agreement with the observed compact shadow and asymmetric brightness distribution, suggesting that dark matter may leave observable imprints on horizon-scale images. Our results position PFDM as a physically motivated extension to the Kerr geometry and highlight a promising astrophysical pathway for probing dark matter near BHs with current and future VLBI campaigns.
			
		\end{abstract}
		
		
		
		\begin{keyword}
			General relativity \sep Black hole \sep Thin accretion disk
			
			
		\end{keyword}
		
	\end{frontmatter}
	
		
		
		\section{Introduction}
		\label{sec:intro}
		\par
		In the standard cosmological framework, the energy composition of the Universe is estimated to consist of approximately 68\% dark energy, 27\% dark matter, and 5\% baryonic matter. Despite the absence of direct detection, extensive observational evidence supports the existence of dark matter. Bahcall \textit{et al.} identified a disk-like distribution of dark matter through a comprehensive survey of K-dwarf stars~\cite{Bahcall:1991qs}. Likewise, Zaritsky \textit{et al.} provided robust confirmation for extended dark halos beyond the optical disks of spiral galaxies by analyzing velocity dispersions of satellite galaxies~\cite{Zaritsky:1996ch}. Mateo's detailed examination of velocity dispersions in dwarf spheroidal galaxies further reinforced the evidence for dark matter~\cite{Mateo:1998wg}. Additionally, Koopmans employed strong gravitational lensing techniques to detect significant dark matter distributions in elliptical galaxies~\cite{Koopmans:2002qh}. Taken together, these observational results have motivated extensive theoretical investigations into the influence of dark matter on the dynamics and physical properties of BHs.

		\par
		BHs serve as unparalleled cosmic laboratories for probing interactions between compact objects and surrounding dark matter-dark energy environments. Kiselev initially proposed modeling dark matter as a perfect fluid~\cite{kiselev2003quintessential,Kiselev:2002dx}, laying important theoretical foundations. Li \textit{et al.} subsequently derived static solutions and spherical approximations applicable to galaxy systems harboring supermassive BHs~\cite{Li:2012zx}. Building upon this framework, Xu \textit{et al.} investigated rotating BH solutions embedded in PFDM, addressing more realistic astrophysical scenarios~\cite{Xu:2017bpz}. These developments have stimulated a broad range of subsequent studies of PFDM-influenced BH models~\cite{Rizwan:2018rgs,Narzilloev:2020qtd,Cao:2021dcq,Zhou:2022eft,Rizwan:2023ivp}, significantly enriching our understanding of their role in cosmic structure formation and evolution.

		\par
		A major breakthrough in astrophysical observations was achieved with the horizon-scale imaging of the supermassive black holes M87$^*$ and Sgr A$^*$ by EHT \cite{EventHorizonTelescope:2019dse,EventHorizonTelescope:2022wkp}. While the images of M87$^*$ revealed a prominent relativistic jet alongside the accretion structure \cite{Lu:2023bbn}, Sgr A$^*$ presents as a compact source without signs of a similarly powerful, sustained jet \cite{janssen2021event}. These observations provide unprecedented direct probes into the physics of accretion flows in strong gravitational fields, thereby substantially advancing our understanding of BH observational signatures. Particularly noteworthy developments have occurred in General Relativistic Magnetohydrodynamic (GRMHD) simulations for Kerr BHs~\cite{akiyama2019first,EventHorizonTelescope:2022wok,EventHorizonTelescope:2022urf,ripperda2022black,Yuan:2022mkw,nalewajko2024chaotic}. For instance, Yang \textit{et al.} demonstrated that kink-instability-driven magnetic reconnection coupled with non-thermal electron acceleration processes could successfully reproduce key structural features observed in the M87$^*$ jet~\cite{yang2024modeling}. Nevertheless, discrepancies remain between state-of-the-art GRMHD simulations and EHT observations, notably regarding finer-scale features such as disk asymmetry. These remaining tensions have motivated explorations beyond conventional Kerr BH frameworks, particularly within modified gravity theories that deviate from General Relativity predictions~\cite{akiyama2022first}. Recent theoretical investigations, utilizing simplified accretion models, seek to isolate distinctive observational signatures characteristic of alternative spacetime geometries~\cite{cunha2020lensing,Zeng:2020dco,Ma:2019ybz,Chen:2020qyp,Zhang:2021hit,Gan:2021pwu,Peng:2020wun,Zeng:2021mok,Guo:2021bhr,guo2022quasinormal,Kuang:2022ojj,Hou:2022eev,Huang:2023ilm,Meng:2024puu,Cui:2024wvz,hu2024influences,zhang2024imaging}. In this context, this study examines rotating BHs immersed in PFDM environments, focusing on how such exotic conditions reshape accretion disk imaging signatures.

		\par
		Previous investigations into PFDM spacetimes established crucial foundations for this work. Haroon \textit{et al.} explored the impact of PFDM parameters on BH shadow images perceived by distant observers, extending their analysis to scenarios with a non-zero cosmological constant~\cite{Haroon:2018ryd}. Heydari-Fard \textit{et al.} employed the Novikov-Thorne accretion model to investigate imaging features of optically thick accretion disks around rotating BHs in PFDM backgrounds~\cite{Heydari-Fard:2022xhr}. In contrast, this paper adopts an optically thin accretion flow model consistent with EHT observational conditions~\cite{EventHorizonTelescope:2021btj}, which more accurately captures radiative transfer processes in low-density accretion environments typical of observed supermassive BHs. Through systematic comparisons against EHT observational constraints, particularly disk asymmetry measurements for M87$^*$, our results elucidate the gravitational imprint of dark matter within extreme astrophysical settings.

		\par
		This paper is organized as follows. Section~\ref{sec:2} is devoted to the theoretical analysis of the spacetime. We begin by investigating the ISCO, which provides fundamental constraints on the accretion dynamics and the parameter space of the BH spin $a$ and the PFDM intensity $k$. Within this framework, we employ an effective potential approach to derive the equations of motion for photons, forming the basis for subsequent shadow imaging. In Section~\ref{sec:3}, we detail numerical simulations of thin accretion disks for a range of parameters and observing frequencies, considering both prograde and retrograde configurations. We then incorporate angular diameter measurements from the Event Horizon Telescope for M87$^{*}$ and Sgr~A$^{*}$ to further constrain the PFDM parameter space. The section culminates in direct comparisons between our synthetic images and EHT observations of M87$^{*}$, allowing a quantitative assessment of how dark matter influences the BH shadow and disk emission structure. Finally, in section~\ref{sec:4}, we summarize our results.

		\section{Relativistic Orbital Dynamics: Shadows, Isco, and Effective Potentials}
		\label{sec:2}
		\par
		The interaction between dark matter and the gravitational field is introduced through minimal coupling, described by the action \cite{kiselev2003quintessential,Li:2012zx}
		\begin{eqnarray}
			\label{eq:1}
			\mathcal{S} = \int \mathrm{d}^4 x \sqrt{-g} \left( \frac{R}{16\pi} + \mathcal{L}_{\rm DM} \right),
		\end{eqnarray}
		where $\mathcal{L}_{\rm DM}$ denotes the Lagrangian density of dark matter, considered to be non-interacting with ordinary matter. The corresponding Einstein field equations are
		\begin{equation}
			\label{eq:2}
			R_{\mu\nu} - \frac{1}{2} g_{\mu\nu} R = 8\pi T_{\mu\nu}^{T},
		\end{equation}
		where $ T_{\mu \nu}^{T}= T_{\mu \nu}^{M} + T_{\mu \nu}^{DM}$ is the total energy-momentum tensor, which accounts for the interaction between ordinary matter and dark matter. For a static and spherically symmetric distribution of PFDM, BH solution modifies the effective mass function into the form \cite{kiselev2003quintessential,Li:2012zx}
		\begin{equation}
			\label{eq:3}
			m(r) = M - \frac{k}{2} \ln \left( \frac{r}{|k|} \right),
		\end{equation}
		where $M$ is the bare mass of BH. The parameter $k$ appears naturally as an integration constant when solving Einstein's equations with a PFDM source, and thus characterizes the overall intensity of the surrounding dark matter halo.
		
		The rotating PFDM BH is constructed via the Newman-Janis algorithm \cite{Xu:2017bpz}, yielding the Kerr-PFDM line element
		\begin{eqnarray}
			\label{eq:4}
			{\rm d}s^{2} & = & -\left(1-\frac{2m(r)r}{\Sigma}\right){\rm d}t^{2} + \frac{\Sigma}{\Delta}{\rm d}r^{2} + \Sigma{\rm d}\theta^{2} \nonumber\\
			& & - \frac{4a m(r)r\sin^{2}\theta}{\Sigma}{\rm d}t{\rm d}\phi + \left[r^{2}+a^{2}+\frac{2a^{2} m(r)r\sin^{2}\theta}{\Sigma}\right]\sin^{2}\theta {\rm d}\phi^{2},
		\end{eqnarray}
		with 
		\begin{eqnarray}
			\label{eq:5}
			\Sigma & = & r^{2}+a^{2}\cos^{2}\theta, \\
			\label{eq:6}
			\Delta & = & r^{2}+a^{2}-2m(r)r.
		\end{eqnarray}
		The parameter $k$ establishes a direct connection to dark matter through the energy-momentum tensor. In an orthonormal basis, the stress-energy tensor of PFDM is diagonal \cite{Xu:2017bpz},
		\begin{equation}
			\label{eq:7}
			T^{\mu}_{\ \nu} = {\rm diag}(-\rho, P_{r}, P_{\theta}, P_{\phi}),
		\end{equation}
		with components
		\begin{eqnarray}
			\label{eq:8}
			\rho & = & \frac{k r}{8\pi (r^{2} + a^{2} \cos^{2}\theta)^{2}}, \\
			\label{eq:9}
			P_{r} & = & -\rho, \\
			\label{eq:10}
			P_{\theta} & = & P_{\phi} = -P_{r} - \frac{k}{16\pi r (r^{2} + a^{2} \cos^{2}\theta)}.
		\end{eqnarray}
		This proportionality demonstrates that $k$ directly sets the overall scale of the dark matter energy density: larger values of $|k|$ correspond to a denser PFDM distribution around the BH. The energy density $\rho$ is linearly proportional to $k$, establishing $k$ as a genuine dark matter intensity parameter rather than merely a mathematical constant. Moreover, the sign of $k$ determines the gravitational nature of PFDM: $k > 0$ corresponds to an attractive effect of dark matter on the spacetime around BH, while $k < 0$ leads to a repulsive effect, similar to some dark energy models \cite{Xu:2017bpz}. Therefore, in the present work we focus on the case $k > 0$ to study the effects of attractive dark matter. In the limit of vanishing dark matter ($k = 0$), the spacetime correctly reduces to the vacuum Kerr solution, as $m(r) \to M$ and the energy-momentum tensor vanishes. Only when both $k=0$ and $a=0$ does it further reduce to the Schwarzschild metric. Throughout this work, we adopt geometrized units ($G=c=1$), in which $k$ has dimensions of length (or equivalently, mass), and thus serves as a well-defined intensity parameter for PFDM in both static and rotating BH configurations. Consequently, any observational constraint on $k$ can be straightforwardly translated into a bound on the PFDM density profile in the galactic nucleus. 
		
		The BH event horizon radius can be obtained by solving the equation $\Delta=0$. Under the PFDM framework, this condition may yield non-physical solutions corresponding to naked singularities when no real horizon exists. Figure 1 illustrates this dichotomy, where the green parameter space corresponds to BH configurations, while the white region exclusively supports naked singularities, thereby clearly delineating the constraints on admissible parameter combinations. Furthermore, within the physically allowed parameter space where BH solutions exist, the permissible range of the spin parameter $a$ varies non-monotonically with PFDM intensity $k$: it first contracts and then expands as $k$ increases. This behavior highlights the non-trivial interplay between PFDM parameters and BH spacetime properties, underscoring the necessity of a careful parameter selection to ensure physical consistency.
		\begin{center}
			\includegraphics[width=8.5cm,height=6.5cm]{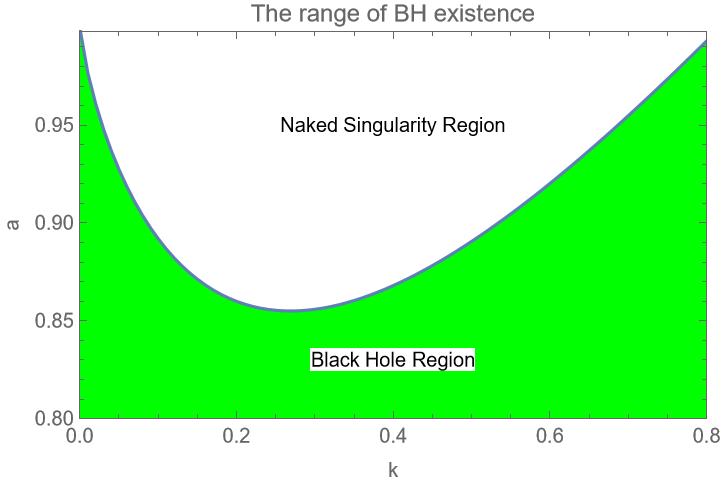}
			\parbox[c]{15.0cm}{\footnotesize{\bf Fig~1.}  
				Parameter space of BH solutions in rotating spacetime with PFDM. The green region marks the solutions that satisfies the horizon condition $\Delta(r) = 0$, and the white region marks naked singularity  that violates this condition.}
			\label{fig1}
		\end{center}

		\subsection{Shadow}
		\label{sec:2-1}
		\par
		The size of BH shadow and the shape of the accretion disk highly depend on the properties of BH. We employ the Hamilton-Jacobi equation and the Carter constant separable method to derive the complete geodesic equation for studying the optical appearance of BH and the motion of photons around it. The specific form of the Hamilton-Jacobi equation is \cite{carter1968global,chandrasekhar1998mathematical}
		\begin{equation}
			\label{eq:11}
			2\frac{{\partial S}}{{\rm d}\sigma}  = -g^{\mu \nu}  \frac{\partial S}{\partial x^{\mu}}\frac{\partial S}{\partial x^{\nu}},
		\end{equation}
		where $\sigma$ is the affine parameter, and $S$ is the Jacobi action, which is given by
		\begin{equation}
			\label{eq:12}
			S= S_{r}(r)+S_{\theta}(\theta) + L \phi-E t.
		\end{equation}
		
		The characteristics of geodesics around a rotating BH in PFDM are completely determined by the energy $E$, angular momentum $L$, and the Carter constant $\mathcal{C}$. The four-momentum $p^{\rm \mu}$ of photons moving along their trajectories can be expressed as \cite{chandrasekhar1998mathematical}
		\begin{eqnarray}
			\label{eq:13}
			&\frac{\Sigma}{{\rm d}\sigma}p^{\rm r} = \pm_{\rm r} \sqrt{\mathcal{A}(r)}, \\
			\label{eq:14}
			&\frac{\Sigma}{{\rm d}\sigma}p^{\rm \theta} = \pm_{\rm \theta} \sqrt{\mathcal{B}(\theta)}, \\
			\label{eq:15}
			&\frac{\Sigma}{{\rm d}\sigma}p^{\rm \phi} =\frac{a}{\Delta}(E r^{2}+E a^{2}-L a)+\frac{L}{\sin^{2}\theta}-E a, \\
			\label{eq:16}
			&\frac{\Sigma}{{\rm d}\sigma}p^{\rm t} = \frac{a^{2}+r^{2}}{\Delta(r)}(E a^{2}+E r^{2}-L a)+L a-E a^{2}\sin^{2}\theta,
		\end{eqnarray}
		where the symbols $\pm_{\rm r}$ and $\pm_{\rm \theta}$ represent the signs of $p^{\rm r}$ and $p^{\rm \theta}$, respectively. The $\mathcal{A}(r)$ and $\mathcal{B}(\theta)$ are the radial and angular potentials, respectively, and their specific forms are
		\begin{eqnarray}
			\label{eq:17}
			&\mathcal{A}(r) = -\Delta(r) [(E a-L)^{2}+\mathcal{C}]+(E a^{2}+E r^{2}- L a)^{2}, \\
			\label{eq:18}
			&\mathcal{B}(\theta) =\mathcal{C}- L^{2}\cot^{2}\theta + E^{2}a^{2}\cos^{2}\theta.
		\end{eqnarray}
		To elucidate the physical implications of these quantities more clearly, we introduce the following two key impact parameters \cite{carter1968global}:
		\begin{equation}
			\label{eq:19}
			\xi = \frac{L}{E},~~~~~\eta = \frac{\mathcal{C}}{E^{2}}.
		\end{equation}
		By substituting these parameters into Eq. (\ref{eq:17})--(\ref{eq:18}), we can further simplify the relevant expressions. Specifically, for photons located at the radius $r = r_{p}$ (photon orbit radius), the critical impact parameters for their unstable circular orbits can be determined by solving the equation $\mathcal{A} = \frac{\partial \mathcal{A}}{\partial r} = 0$, yielding
		
		\begin{eqnarray}
			\label{eq:20}
			&\xi=\frac{(2f(r)+r f'(r))(a^{2}+r^{2})-4a^{2}-4f(r)r^{2}}{2af(r)+arf'(r)}, \\
			\label{eq:21}
			&\eta=\frac{r^{3}[8a^{2}f'(r)-(-2f(r)+rf'(r))^{2}r]}{(2f(r)+rf'(r))^{2}a^{2}}.
		\end{eqnarray}
		where $f(r)=1-\frac{2M}{r} -\frac{k}{r} \ln \frac{r}{\mid k \mid}$, and $f'(r)$ is the derivative of $f(r)$ over $r$. For an observer at spatial infinity, the BH shadow is projected onto the celestial plane described by coordinates $(\alpha, \beta)$, where $\alpha$ represents the apparent displacement along the axis of symmetry and $\beta$ measures the perpendicular distance. These celestial coordinates are derived from the photon's conserved quantities through asymptotic matching of the geodesic equations \cite{chandrasekhar1998mathematical}. Specifically, for photons reaching the observer with inclination angle $\theta$, the coordinates relate to the impact parameters as \cite{chandrasekhar1998mathematical}:
		\begin{eqnarray}
			\label{eq:22}
			&\alpha=-\frac{\xi}{\sin \theta}, \\
			\label{eq:23} &\beta=\pm \sqrt{\eta + a^{2}\cos^{2}\theta-\xi^{2}\cot^{2}\theta}.
		\end{eqnarray}

		Figure 2 displays the BH shadow contours on the $\alpha-\beta$ plane across varying parameter sets $\{a, k\}$. For a fixed PFDM intensity $k$, enhancing the spin parameter $a$ progressively distorts the shadow from a quasi-circular profile to a pronounced D-shaped asymmetry. Furthermore, for a fixed spin parameter $a$, the shadow area contracts with increasing PFDM intensity $k$ until a critical value is reached, beyond which it expands—a non-monotonic dependence attributed to dark matter-induced modifications of the spacetime curvature. This behavior reflects the competing influence of spin and PFDM on null geodesics near the BH and will be discussed in more detail in the following sections.
		\begin{center}
			\includegraphics[width=4.5cm,height=4.5cm]{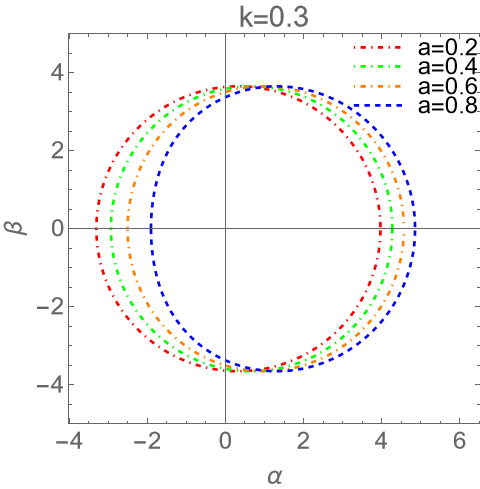}
			\includegraphics[width=4.5cm,height=4.5cm]{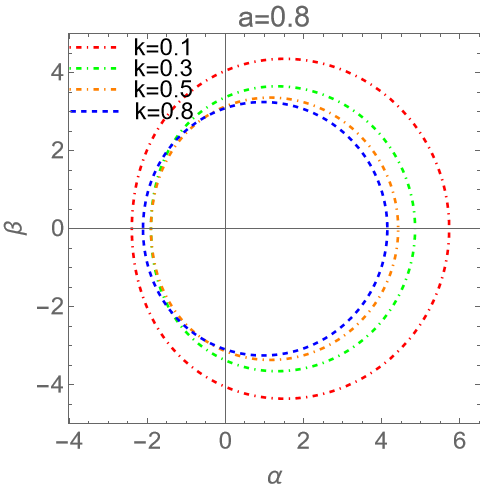}
			\includegraphics[width=4.5cm,height=4.5cm]{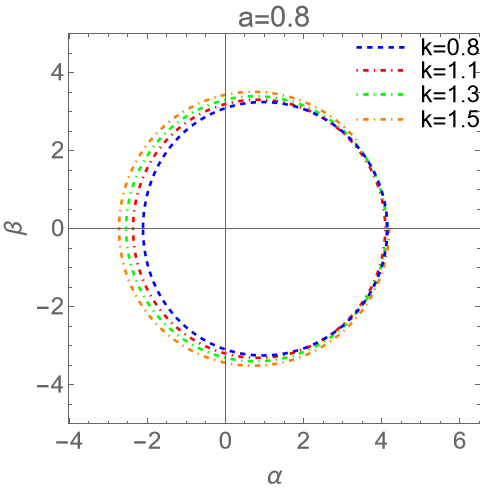}
			\parbox[c]{15.0cm}{\footnotesize{\bf Fig~2.}  
				The shadow contours on the $\alpha-\beta$ plane by adopting various parameter sets of $\{a, k\}$. The mass of the BH is set as $M = 1$.}
			\label{fig2}
		\end{center}

		\subsection{ISCO and Effective Equations}
		\label{sec:2-2}
		
		\par
		Previous studies of BH accretion disks typically take the ISCO as the inner boundary. Particles located outside this orbit can engage in stable circular motion around the BH. However, once particles cross the ISCO, their motion becomes unstable, ultimately leading them into a plunging trajectory toward the event horizon, where they will eventually fall into the BH. Therefore, a clear distinction between particle dynamics inside and outside the ISCO is essential for modeling accretion flows.
		
		\begin{center}
			\includegraphics[width=6.5cm,height=5cm]{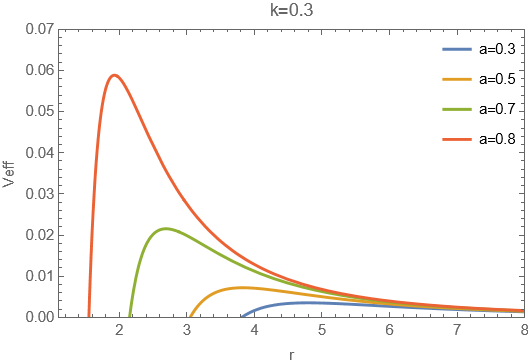}
			\includegraphics[width=6.5cm,height=5cm]{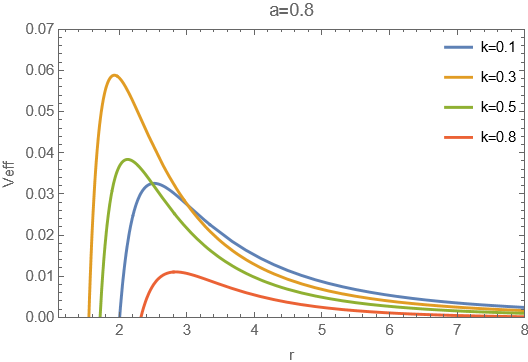}
			\parbox[c]{15.0cm}{\footnotesize{\bf Fig~3.}  
				The BH effective potentials by adopting various parameter sets of $\{a, k\}$ as marked in each panels.}
			\label{fig3}
		\end{center}
		The radial equation governing particle motion adheres to the normalization conditions for energy and angular momentum and can be expressed as \cite{Hou:2022eev}
		\begin{equation}
			\label{eq:24}
			u^{\rm r}= - \sqrt{-\frac{V(r,E,L)}{g^{\rm rr}}}.
		\end{equation}
		where
		\begin{equation}
			\label{eq:25}
			V(r,E,L) = (g^{\rm tt}E^{2} + g^{\rm \phi\phi}L^{2}-2g^{\rm t\phi} EL + 1)|_{\theta=\frac{\pi}{2}},
		\end{equation}
		represents the effective potential of the BH in PFDM. Figure 3 shows the effective potential $V$ as a function of the radial coordinate $r$. It is observed that at fixed PFDM intensity, the peak of the effective potential is significantly reduced and shifts to larger radii as the spin parameter increases. When the spin is held constant, an intriguing trend emerges: with increasing PFDM intensity $k$, the peak initially rises and shifts inward; however, upon a further increase in $k$, it then decreases and shifts back outward. This non-monotonic behavior stems from the competition between the spacetime curvature induced by PFDM and the gravitational pull of BH. Notably, at sufficiently large radii, the effective potential becomes insensitive to variations in $k$, indicating that PFDM primarily modifies the near-horizon spacetime structure.
		
		A detailed analysis of the ISCO is essential for determining the inner boundary of the accretion disk in the PFDM spacetime. The ISCO radius is obtained by imposing the marginal stability conditions for equatorial circular orbits,
		\[
		V(r,E,L)=0, \qquad \partial_r V(r,E,L)=0, \qquad \partial^2_r V(r,E,L) = 0.
		\]
		For particles located outside the ISCO radius ($r>r_{\rm ISCO}$), their motion is governed by the effective potential $V(r,E,L)$. Once the ISCO is crossed, circular motion becomes unstable and the particle enters a plunging trajectory. In this regime ($r<r_{\rm ISCO}$), the radial velocity is described by \cite{Hou:2022eev}
		\begin{equation}
			\label{eq:26}
			u^{r}_{c}= - \sqrt{-\frac{V(r,E_{\rm ISCO},L_{\rm ISCO})}{g^{\rm rr}}} \Bigg|_{\theta=\frac{\pi}{2}} .
		\end{equation}
		
		To investigate the dependence of the ISCO on the PFDM intensity, we fix the black hole spin parameter $a$ and numerically solve the ISCO conditions while varying the PFDM parameter $k$. The solutions are searched within the radial range $r \lesssim 25\,r_g$, which corresponds to the radial extent of the accretion disk adopted as the computational domain in the present model and is consistently used in the subsequent ray-tracing and imaging calculations throughout this work. Within this adopted radial domain, the ISCO is unique and varies smoothly with $k$ for $k \lesssim 0.5$. For larger values of $k$, multiple ISCO solutions can appear within the same radial range, preventing the identification of a single, well-defined inner edge of the accretion disk. Such configurations are therefore incompatible with the steady, geometrically thin disk model and imaging framework assumed in this paper. Accordingly, we restrict our subsequent analysis to $k \le 0.5$.
		\begin{figure}[H] 
			\centering
			\includegraphics[width=6.8cm,height=5cm]{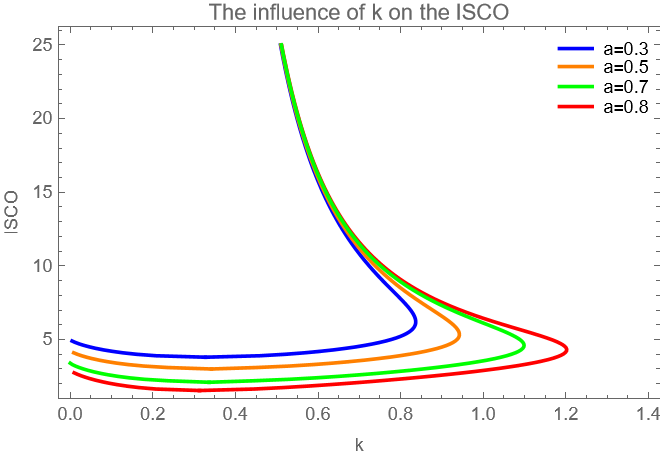}
			\parbox[c]{15.0cm}{\footnotesize{\bf Fig~4.} 
				The ISCO as a function of the PFDM intensity by adopting $a=0.3,0.5, 0.7, 0.8$, respectively. The ISCO solutions are shown within the radial domain adopted in our disk and imaging model ($r \lesssim 25r_g$).}
			\label{fig4}
		\end{figure}	
		For completeness, we note that if $k$ is increased beyond the range considered above, the two ISCO branches eventually merge at a critical value of $k$, which appears as the rightmost turning point in Fig.~4 for a given spin $a$; this solution formally satisfies the ISCO conditions and corresponds to a legitimate marginally stable circular orbit, but it represents a boundary configuration separating the single-ISCO and double-ISCO regimes and exists only at an isolated value of $k$ rather than persisting over a finite parameter interval, and since our analysis focuses on accretion disk configurations that admit a well-defined and continuously varying inner edge within the adopted computational domain, this turning-point ISCO is not included in the subsequent disk modeling and imaging calculations.

		\section{Simulated Appearance and Comparison}
		\label{sec:3}
		
		\par
		To map photon trajectories from the accretion disk to a distant observer, we employ the zero-angular-momentum observer (ZAMO) framework. The ZAMO's orthonormal tetrad formalism---representing a locally non-rotating reference frame with vanishing angular momentum---provides a natural basis for projecting the photon four-momentum. The tetrad vectors are given by \cite{Hu:2020usx}:
		\begin{eqnarray}
			\label{eq:27}
			&e_{\rm t}=\varpi \partial_{t} + \lambda \partial_{\phi},~~~e_{\rm r}=\frac{1}{\sqrt{g_{\rm rr}}}\partial_{r},\\
			\label{eq:28}
			&e_{\rm \theta}=-\frac{1}{\sqrt{g_{\rm \theta\theta}}} \partial_{\theta},~~~~~e_{\rm \phi}=\frac{1}{\sqrt{g_{\rm \phi\phi}} }\partial_{\phi},
		\end{eqnarray}
		where the coefficients read
		\begin{equation}
			\label{eq:29}
			\varpi = \sqrt{\frac{g_{\rm \phi\phi}}{g^{2}_{\rm t\phi}- g_{\rm tt}g_{\rm \phi\phi}}},~~~\lambda=-\frac{g_{\rm t\phi}}{g_{\rm \phi\phi}}\sqrt{\frac{g_{\rm \phi\phi}}{g^{2}_{\rm t\phi}- g_{\rm tt}g_{\rm \phi\phi}}}.
		\end{equation}
		In this ZAMO frame, the components of a photon's four-momentum are expressed as \cite{wang2024image}
		\begin{eqnarray}
			\label{eq:30}
			&P^{\rm t}=E(\varpi - \lambda \xi),~~~~~P^{\rm r}= \frac{E}{\sqrt{g_{\rm rr}}} \frac{\pm_{\rm r} \sqrt{\mathcal{A}(r)}}{\Delta_{\rm r}},\\
			\label{eq:31}
			&P^{\rm \theta} =  \frac{E}{\sqrt{g_{\rm \theta\theta}}} \frac{\pm_{\rm \theta} \sqrt{\mathcal{B}(\theta)}}{\Delta_{\rm \theta}},~~~~~P^{\phi}=  \frac{E \xi}{\sqrt{g_{\rm \phi\phi}}}.
		\end{eqnarray}
		To connect these momentum components with the observer's image, we employ a fisheye camera model based on stereographic projection, which is ideal for capturing wide-field images of BH shadow. The observer is positioned at point $O$ with coordinates $(0, r_{\rm o}, \theta_{\rm o}, 0)$, while the center of the projection screen is denoted as point $M$. For any arbitrary pixel point $P$ on the screen, the incoming photon direction is characterized by the celestial coordinates $(\Theta, \Psi)$, where $\Theta$ is the angle between the light ray and the optical axis $OM$, and $\Psi$ is the azimuthal angle. In this framework, the screen coordinates $(x,y)$ are defined in angular units through a stereographic projection \cite{Hu:2020usx}, and are related to $(\Theta,\Psi)$ by
		\begin{equation}
			\label{eq:32}
			x = -2 \tan \frac{\Theta}{2} \sin\Psi, \quad y = -2 \tan \frac{\Theta}{2} \cos\Psi.
		\end{equation}
		Here, the coordinates $(x,y)$ represent dimensionless angular positions on the observer’s image plane, corresponding to a normalization by the observer distance. In our setup, the observer is placed at asymptotically large radius, so that any overall multiplicative scale associated with the observer distance would only rescale the image uniformly and does not affect the image morphology or the relative structure of the photon rings.
		
		The field-of-view angle $\alpha_{\rm fov}$ defines the angular extent of the imaging system, and for a unit distance between $O$ and $M$, the physical screen length is $L = 2 \tan(\alpha_{\rm fov}/2)$. Since light propagation is reversible, we initialize the numerical integration at the observer's location $O$. The celestial coordinates $(\Theta, \Psi)$ for each pixel determine the initial photon momentum components via Eqs.~(\ref{eq:30})--(\ref{eq:31}). By solving the Hamiltonian equations for null geodesics, we trace the complete photon trajectories backward in time. This backward ray-tracing approach allows us to determine whether each photon originates from the accretion disk or falls into BH, thereby reconstructing the shadow structure and the surrounding emission features.
		
		\par
		We applied this fisheye camera to capture the shadows of BH in PFDM under various conditions, as shown in Figure 5. To facilitate visualization, the screen is divided into four quadrants, which improves the clarity of photon trajectories in different regions. The results show that a high spin significantly distorts the light around the BH, resulting in a distinctly D-shaped shadow. Moreover, increasing the PFDM intensity from 0.1 to 0.5 makes the Einstein ring shift inward significantly and the degree of light distortion surrounding the BH diminish. These trends are consistent with the behavior inferred from the effective potential analysis, underscoring the intricate interplay between the BH spin, PFDM intensity, and the underlying spacetime geometry.
		
		\begin{center}
			\includegraphics[width=4.5cm,height=4.5cm]{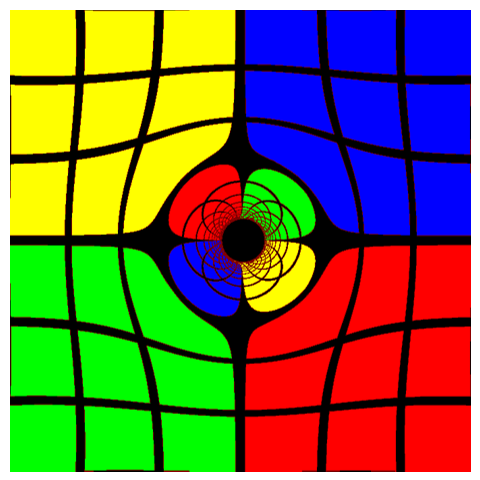}
			\includegraphics[width=4.5cm,height=4.5cm]{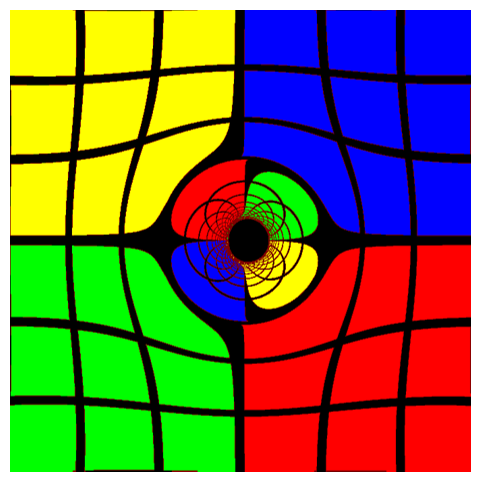}
			\includegraphics[width=4.5cm,height=4.5cm]{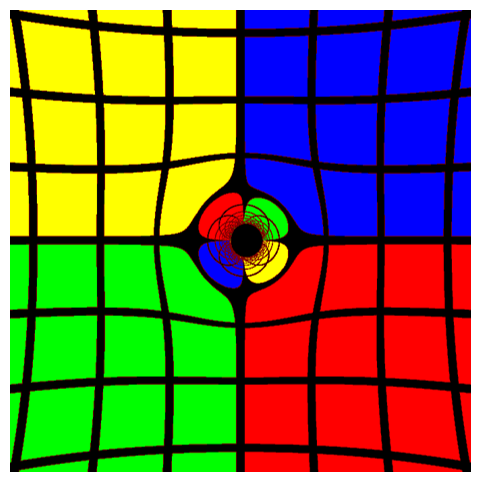}
			\parbox[c]{15.0cm}{\footnotesize{\bf Fig~5.}  
				Images of the black hole shadow in the PFDM model obtained via numerical ray tracing for varying parameters. From left to right, the parameters are: (Left) spin $a=0.3$, PFDM intensity $k=0.1$; (Middle) $a=0.8$, $k=0.1$; (Right) $a=0.8$, $k=0.5$. The black hole mass is normalized to $M=1$.}
			\label{fig5}
		\end{center}
		
		\subsection{Accretion Disks}
		\label{sec:3-1}
		\par
		To model observable emission features, interactions between photons and the accretion disk must be taken into account once rays intersect the disk plane. For simplicity, we assume that the refractive effects of the accretion disk medium can be neglected. The intensity of the BH accretion disk as observed by an observer in a locally non-rotating (ZAMO) frame is given by \cite{Lindquist:1966igj}
		
		\begin{equation}
			\label{eq:33}
			\frac{{\rm d}}{{\rm d}\lambda} \left( \frac{I_{\rm \nu}}{\nu^{3}} \right) = \frac{J_{\rm \nu} - \kappa_{\rm \nu} I_{\rm \nu}}{\nu^{2}},
		\end{equation}
		where $\lambda$ represents the affine parameter, and $I_{\rm \nu}$, $J_{\rm \nu}$, and $\kappa_{\rm \nu}$ are the specific intensity, emissivity, and absorption coefficient at frequency $\nu$, respectively. We consider the propagation of light in a vacuum, where both $J_{\rm \nu}$ and $\kappa_{\rm \nu}$ are assumed to be zero. As a result, the quantity $I_{\rm \nu}/\nu^{3}$ remains conserved along the geodesic path.
		
		The adopted accretion disk model is geometrically thin and optically thin, ensuring that the emission and absorption coefficients remain constant along photon trajectories intersecting the disk. Integrating Eq.~\ref{eq:33} along each ray, the observed intensity on the distant screen can be written as \cite{Hou:2022eev}
		\begin{equation}
			\label{eq:34}
			I_{\nu_o} = \sum_{s=1}^{N_{\rm max}} 
			\left( \frac{\nu_o}{\nu_s} \right)^3 
			\frac{J_s}{\tau_{s-1}} 
			\left[ \frac{1 - e^{-\kappa_s f_s}}{\kappa_s} \right],
		\end{equation}
		where $s$ labels the number of equatorial plane crossings of a photon trajectory, $\nu_o$ is the frequency measured by the distant observer, and $\nu_s$ represents the frequency observed in a local stationary frame with the accretion disk in motion. The ratio $g \equiv \nu_o / \nu_s$ defines the redshift factor, whose explicit expression will be discussed in detail below. The quantity $\kappa_s$ denotes the absorption coefficient evaluated at the $s$-th equatorial crossing, while $f_s$ is a ``fudge factor'' introduced to regulate the relative brightness of higher-order photon rings \cite{Gralla:2019drh}.
		
		In general, the cumulative absorption experienced by a photon propagating through a medium is characterized by the optical depth $\tau = \int \kappa_\nu \, dl$, while the corresponding attenuation of the radiation is described by the transmission factor $e^{-\tau}$ \cite{rybicki2024radiative}. In the present geometrically thin disk model, radiative interactions are localized to discrete crossings of the equatorial plane. Accordingly, the cumulative attenuation accumulated prior to the $s$-th crossing is encoded through the factor $\tau_{s-1}$ in Eq.~\ref{eq:34}, which represents the total transmission along the photon trajectory up to that point.
		
		In the optically thin limit considered here, the absorption at each crossing is weak, satisfying $\kappa_s f_s \ll 1$. As a result, the absorption term can be approximated to first order as $1 - e^{-\kappa_s f_s} \approx \kappa_s f_s$, while the accumulated attenuation remains negligible, yielding $\tau_{s-1} \simeq 1$. Under these approximations, Eq.~\ref{eq:34} simplifies to
		\begin{equation}
			\label{eq:35}
			I_{\nu_o} = \sum_{s=1}^{N_{\rm max}} f_s\, g^{3}(r_s)\, J_{\rm model}(r_s),
		\end{equation}
		where $J_{\rm model}(r_s)$ denotes the equatorial emissivity evaluated at the radius $r_s$ corresponding to the $s$-th equatorial crossing. Following the approach established in Refs.~\cite{Chael:2021rjo}, we adopt $f_s = 1.5$ and model the equatorial emissivity profile using a second-order polynomial in logarithmic space: 
		\begin{equation}
			\label{eq:36}
			\log[J_{\rm model}(r)] = A \log[r/r_{\rm H}] + B (\log[r/r_{\rm H}])^{2},
		\end{equation}
		where the two free parameters $A$ and $B$ are directly related to the observation frequency. For the observations of M87$^{*}$ and SgrA$^{*}$ by EHT at a frequency of $230$ GHz, the values of these parameters are set to $A=-2$ and $B=-1/2$. When the observation frequency is decreased to $86$ GHz, these parameters are adjusted to $A=0$ and $B=-3/4$ \cite{EventHorizonTelescope:2019jan}.
		
		The expression for the redshift factor of Eq.~\ref{eq:34} not only effectively describes regions outside the ISCO but also performs well in areas close to the BH event horizon. Therefore, we provide its more detailed expression here. Energy and frequency are directly related by the Planck relation $E=h \nu$, where $h$ is Planck's constant and $\nu$ is the frequency of the photon. As energy and frequency are proportional, the ratio of energies between the observed and emitted photons is equivalent to the ratio of their frequencies, which is expressed as
		\begin{equation}
			\label{eq:37}
			g = \frac{\nu_{\rm o}}{\nu_{\rm s}} = \frac{E}{E_{\rm s}}.
		\end{equation}
		In the context of a BH, particularly near the accretion disk, the energy of a photon can be calculated using its four-momentum and the observer's four-velocity in a local static coordinate system. This gives the energy of the photon as \cite{Gralla:2019drh,cunningham1975effects}
		\begin{equation}
			\label{eq:38}
			E_{s}=p^{(t)}=-p_{\mu}u^{\mu},
		\end{equation}
		where $p_{\mu}$ is the photon's four-momentum and $u^{\mu}=e^{\mu}_{t}$ is the four-velocity of the emitting source. Outside the ISCO, the accretion flow continues to move along a circular orbit with an angular velocity defined by $\Omega_{\rm m}(r) = (\mu^{\rm \varphi}/\mu^{\rm t})_{\mid {r=r_{\rm m}}}$. Thus, the expression for the redshift factor can be rewritten as
		\begin{equation}
			\label{eq:39}
			g_{\rm m} = \frac{\varpi(\xi\frac{g_{\rm t \phi}}{g_{\phi \phi}}+1)}{\zeta(1 - \xi \Omega_{\rm m})},
		\end{equation}
		where $\zeta=\sqrt{{-1}/(g_{tt}+g_{\phi \phi}\Omega_m^{2}+2g_{t \phi}\Omega_{m})}$. For the region inside the ISCO, the accretion flow has a radial velocity $u$ and moves along a plunging trajectory. This leads a redshift factor of
		\begin{equation}
			\label{eq:40}
			g_{\rm m} = \frac{\varpi(\xi\frac{g_{\rm t \phi}}{g_{\phi \phi}}+1)}{\frac{u_{\rm c}^{\rm r} k_{\rm r}}{-k_{\rm t}} + g^{\rm tt} E_{\rm ISCO} - \xi g^{\rm t\phi} E_{\rm ISCO} - g^{\rm t\phi} L_{\rm ISCO} + \xi g^{\rm \phi\phi} L_{\rm ISCO}} .
		\end{equation}
		\par
		We reconstruct images of a perfect fluid-surrounded BH with a thin accretion disk by numerically solving Eqs.~\ref{eq:39}--\ref{eq:40} using a fisheye camera-based ray-tracing scheme. Our analysis first addresses the relativistic Doppler effects arising from the disk's unidirectional rotation. As the observer's inclination angle increases, Doppler boosting gradually dominates over gravitational redshift, creating a prominent blueshifted region on the approaching side of the disk. The location of the maximum blueshift thus serves as a sensitive indicator of the disk's kinematic state. As a representative case, Fig.6 shows results for a BH with spin $a = 0.8$ and perfect fluid intensity $k = 0.5$, viewed at inclination $\theta_0 = 17^\circ$. The background color map classifies photon trajectories by their number of equatorial plane crossings: black indicates the shadow region; light blue corresponds to photons deflected without crossing; pale green and red denote trajectories crossing once or multiple times, respectively. Superposed on this map are circles marking the maximum blueshift locations under different parameter combinations: pink ($a=0.3$, $k=0.1$), yellow ($a=0.8$, $k=0.1$), and blue ($a=0.8$, $k=0.5$). Variations in $a$ and $k$ modify both photon geodesics and the gravitational redshift–Doppler balance, leading to a measurable displacement of the blueshift maximum.
		\begin{center}
			\includegraphics[width=6.5cm,height=6.5cm]{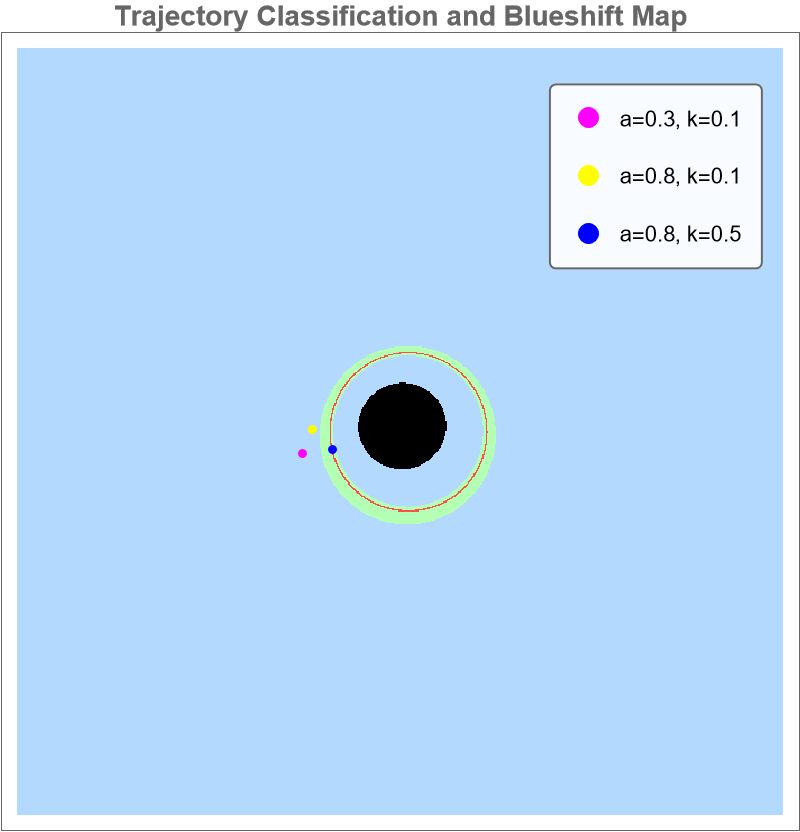}
			\parbox[c]{15.0cm}{\footnotesize{\bf Fig~6.}  
				Photon trajectory classification for a black hole surrounded by perfect fluid, with background computed at $a = 0.8$, $k = 0.5$, $\theta_0 = 17^\circ$. Colors indicate equatorial crossing counts: black (shadow), light blue (deflected), pale green (one crossing), red (multiple crossings). Superposed circles mark maximum blueshift positions for different parameters: pink ($a=0.3$, $k=0.1$), yellow ($a=0.8$, $k=0.1$), blue ($a=0.8$, $k=0.5$).}
			\label{fig6}
		\end{center}
		
		\begin{center}
			\includegraphics[width=12cm,height=12cm]{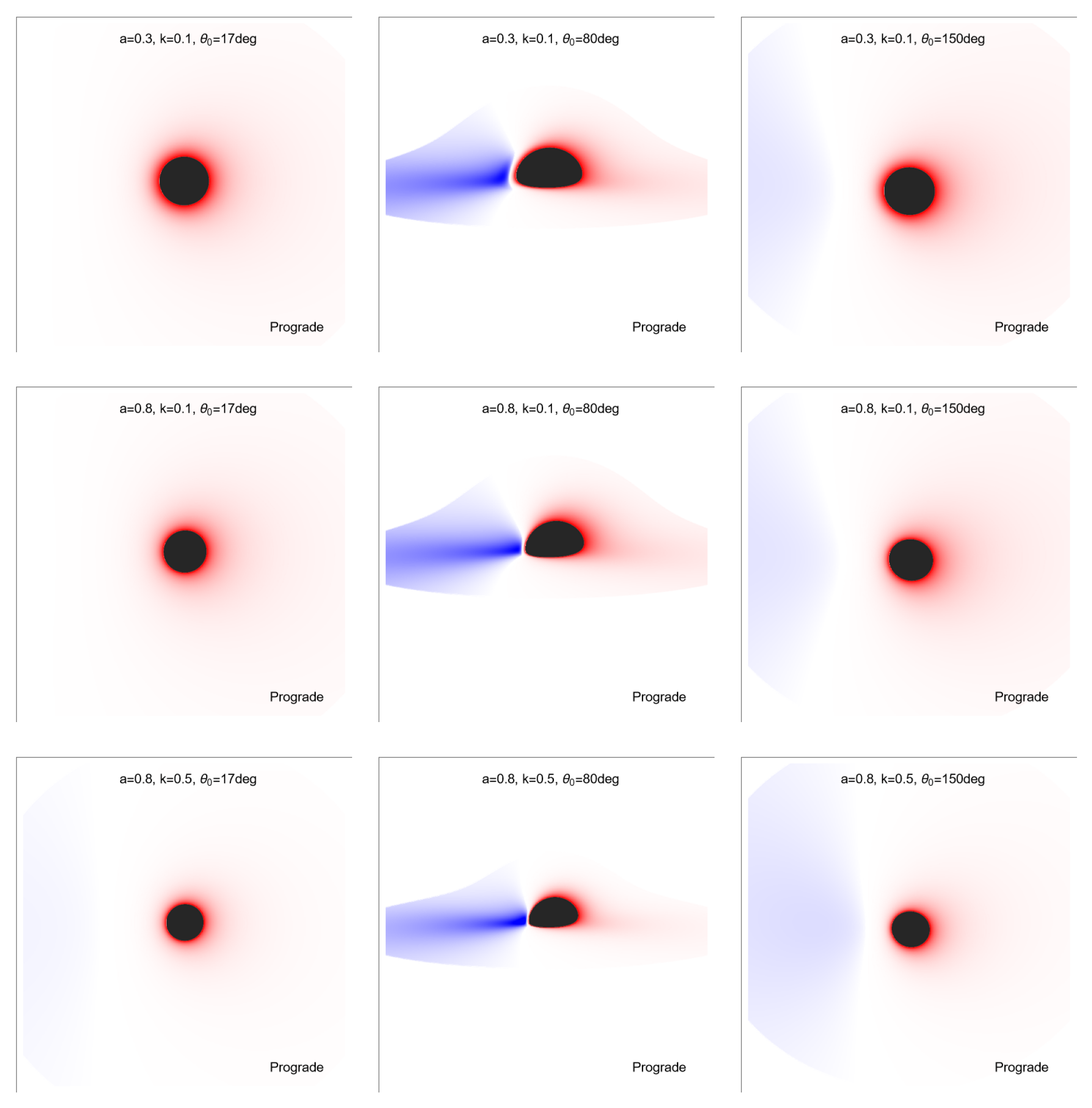}
			\includegraphics[width=1.2cm,height=11.5cm]{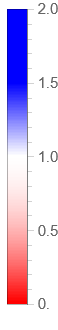}
			\parbox[c]{15.0cm}{\footnotesize{\bf Fig~7.}  
				Redshift ($g<1$, cooler colors) and blueshift ($g>1$, hotter colors) maps for the direct image of a prograde accretion disk around a BH ($M=1$) in PFDM, observed at 230 GHz. The maps are shown for various observer inclination angles $\theta_0$. The panels correspond to different parameter sets $(a, k)$: (Top) $(0.3, 0.1)$; (Middle) $(0.8, 0.1)$; (Bottom) $(0.8, 0.5)$, where $a$ is the spin parameter and $k$ is the PFDM intensity parameter.}
			\label{fig7}
		\end{center}
		\par
		Figure 7 and 8 systematically present the redshift and blueshift distributions for direct and secondary images across various parameter combinations. Key findings for the prograde configuration include: (1) Near-edge-on viewing angles ($\theta_{0} \rightarrow 90^{\circ}$) induce pronounced hemispheric asymmetry due to blueshift enhancement on the approaching side. (2) As the inclination extends beyond 90$^\circ$ ($\theta_{0} \rightarrow 180^{\circ}$), the blueshift progressively diminishes as the observer moves toward a face-on view of the receding side. (3) Parameter dependencies emerge through the distinct modulation patterns of the spin $a$ and PFDM intensity $k$: higher values of $a$ and $k$ lead to a stronger maximum blueshift for prograde accretion. For completeness, we also provide the quantitative results for retrograde accretion flows, which are summarized together with the prograde cases in Table 1. These comparisons highlight the non-trivial coupling between spacetime frame-dragging and the surrounding dark matter halo in shaping the observed relativistic emission patterns.
		\begin{center}
			\includegraphics[width=12cm,height=12cm]{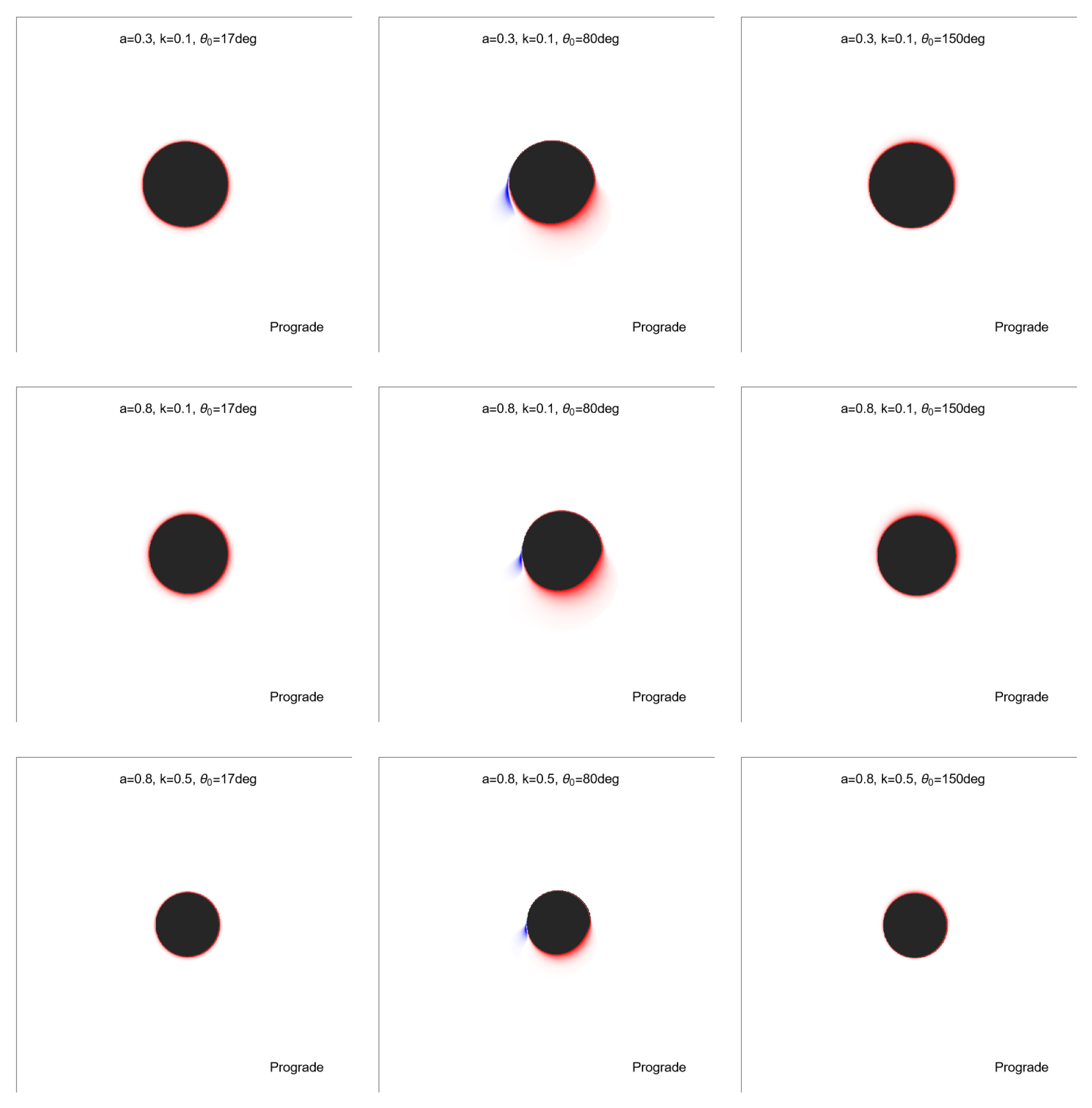}
			\includegraphics[width=1.2cm,height=11.5cm]{hybar.png}
			\parbox[c]{15.0cm}{\footnotesize{\bf Fig~8.}  
				The same as Fig~6 but for the secondary image of the accretion disk.}
			\label{fig8}
		\end{center}

		\begin{center}
			\includegraphics[width=12cm,height=12cm]{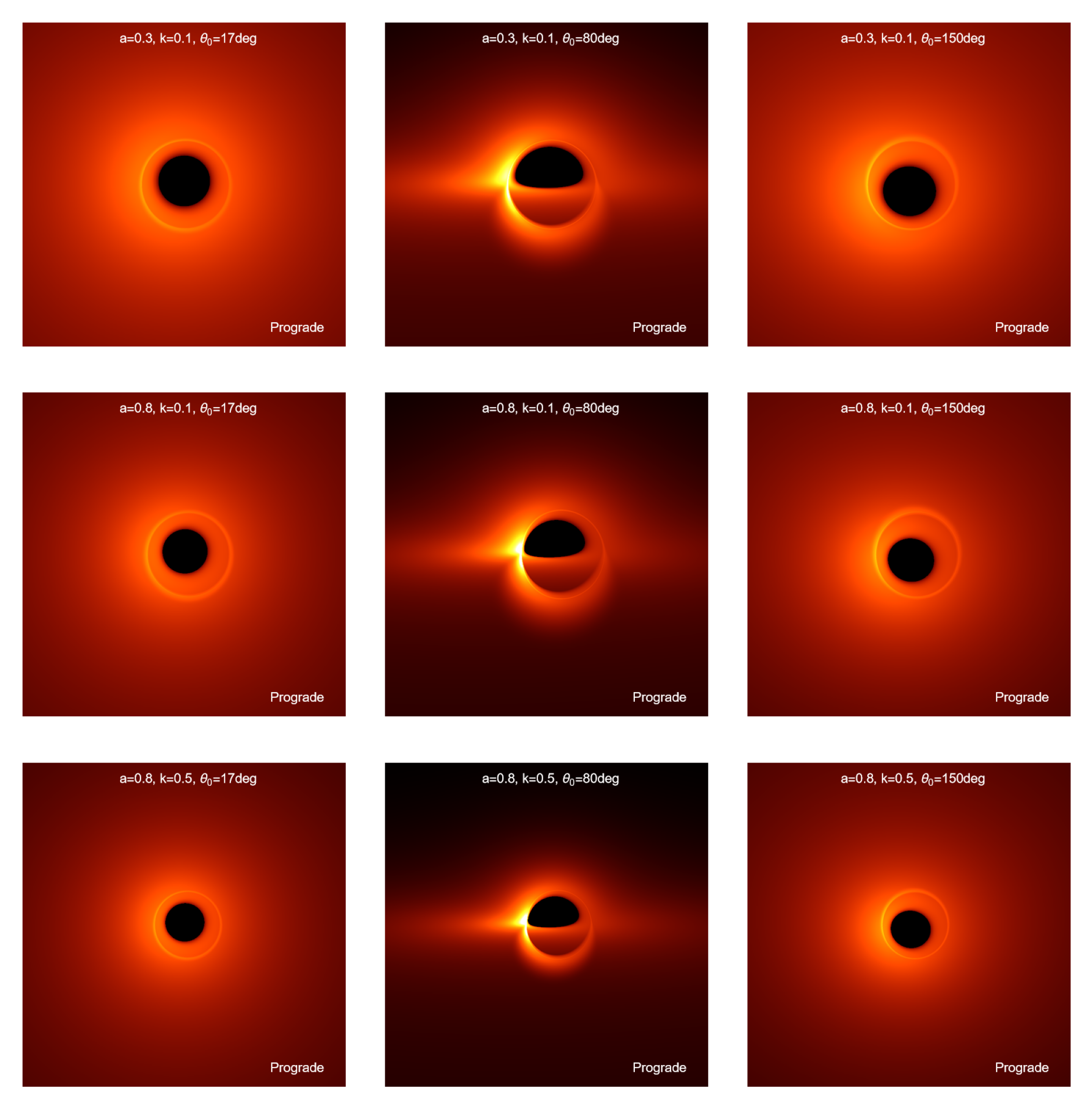}
			\includegraphics[width=1.2cm,height=11.5cm]{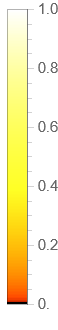}
			\parbox[c]{15.0cm}{\footnotesize{\bf Fig~9.}  
				230 GHz imgaes of a BH ($M=1$) with a prograde accretion disk viewing at different degrees, where the parameter set $(a, k)$ are: (Top) $(0.3, 0.1)$; (Middle) $(0.8, 0.1)$; (Bottom) $(0.8, 0.5)$.}
			\label{fig9}
		\end{center}
		
		\begin{table}[tbp]
			\caption{Maximum Blueshift($g_{\rm max}$) of Direct Image at 230 GHz}
			\label{Tab:1}
			\centering
			\small
			\begin{tabular}{lcccccc}
				\hline
				& \multicolumn{3}{c}{Prograde} & \multicolumn{3}{c}{Retrograde} \\
				\cline{2-4} \cline{5-7}
				$\theta_0$ & $a=0.3$ & $a=0.8$ & $a=0.8$ & $a=0.3$ & $a=0.8$ & $a=0.8$ \\
				& $k=0.1$ & $k=0.1$ & $k=0.5$ & $k=0.1$ & $k=0.1$ & $k=0.5$ \\
				\hline
				$17^\circ$ & 1.01893 & 1.02048 & 1.02247 & 1.01776 & 1.01703 & 1.01757 \\
				$80^\circ$ & 1.60662 & 1.70944 & 1.86558 & 1.54317 & 1.50787 & 1.51369 \\
				$150^\circ$ & 1.05859 & 1.05573 & 1.06811 & 1.05450 & 1.05145 & 1.05308 \\
				\hline
			\end{tabular}
		\end{table}
		
		\begin{center}
			\includegraphics[width=4.4cm,height=3.7cm]{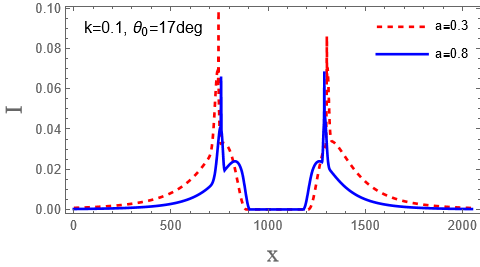}
			\includegraphics[width=4.4cm,height=3.7cm]{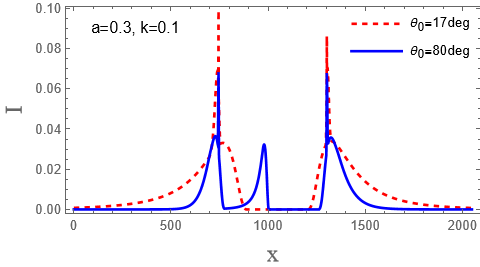}
			\includegraphics[width=4.4cm,height=3.7cm]{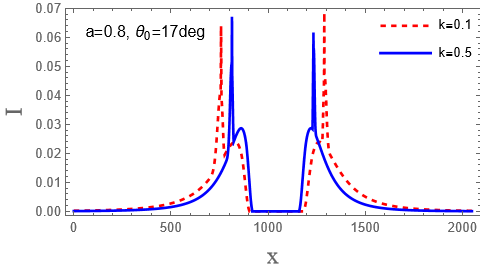}
			\includegraphics[width=4.5cm,height=3.7cm]{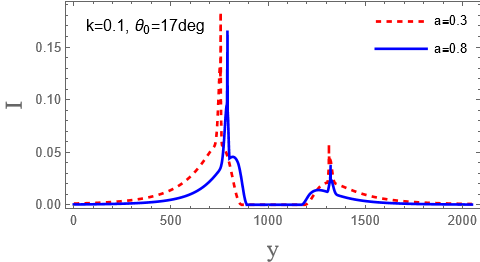}
			\includegraphics[width=4.5cm,height=3.7cm]{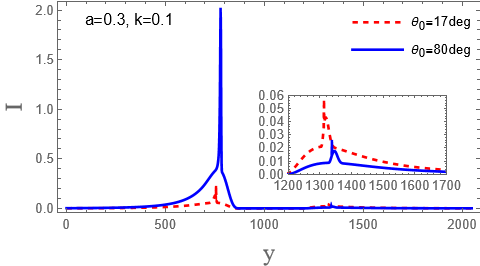}
			\includegraphics[width=4.5cm,height=3.7cm]{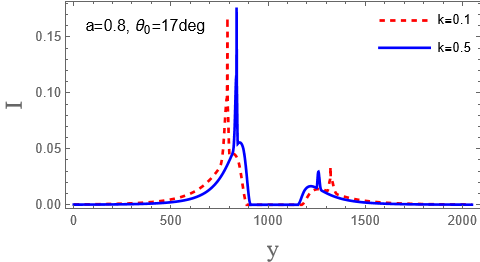}
			\parbox[c]{15.0cm}{\footnotesize{\bf Fig~10.}  
				Intensity profiles for the prograde accretion disk images at 230 GHz shown in Fig~9. Top row: Intensity distribution along the horizontal (x) axis. Bottom row: Intensity distribution along the vertical (y) axis. The columns correspond to the parameter sets $(a, k) = (0.3, 0.1)$ (left), $(0.8, 0.1)$ (middle), and $(0.8, 0.5)$ (right), matching the panels in Fig~9.}
			\label{fig10}
		\end{center}
		
		\begin{center}
			\includegraphics[width=12cm,height=12cm]{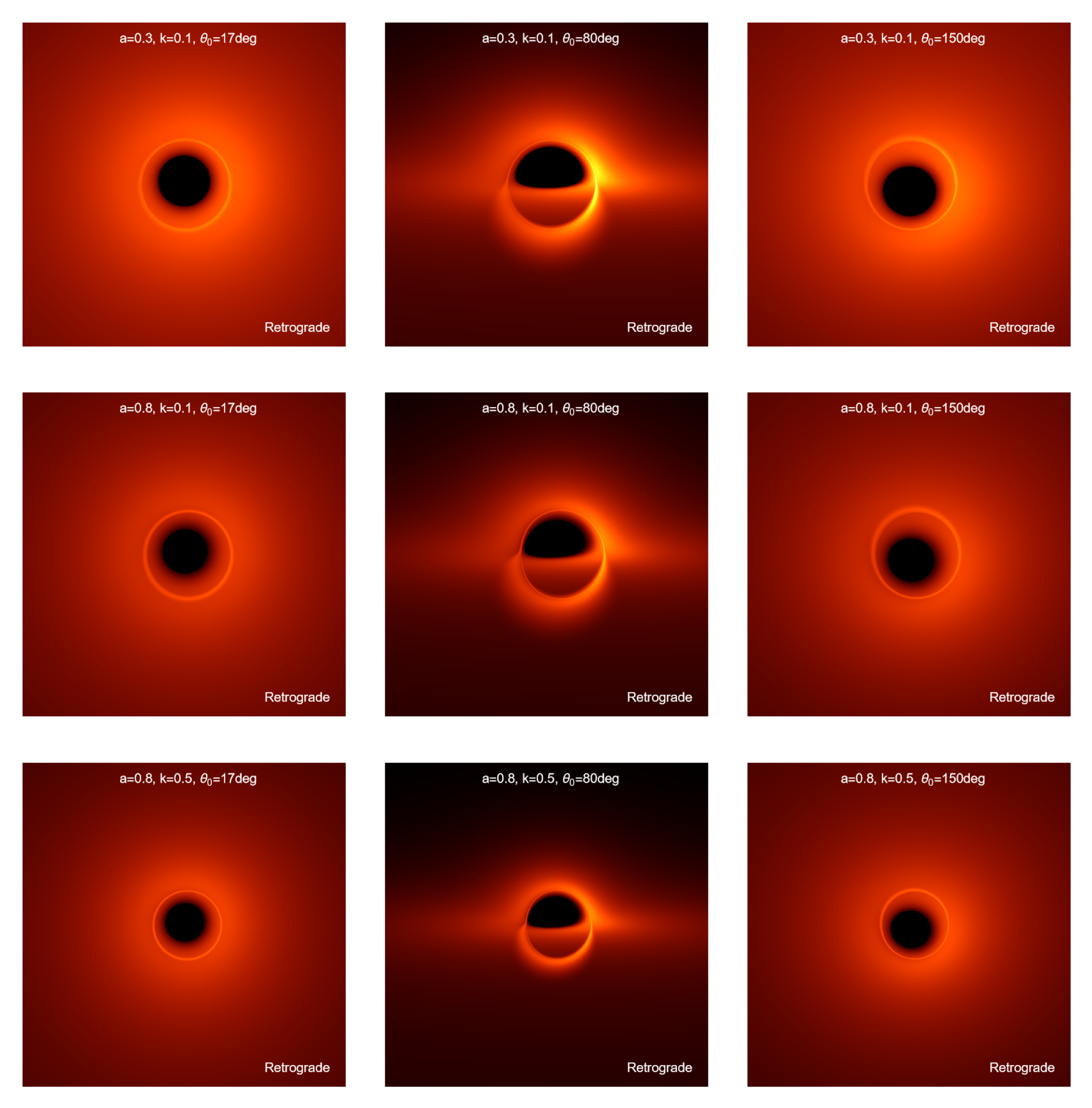}
			\includegraphics[width=1.2cm,height=11.5cm]{fbar.png}
			\parbox[c]{15.0cm}{\footnotesize{\bf Fig~11.}  
				The same as Fig~9 but for the BH with a retrograde accretion.}
			\label{fig11}
		\end{center}

		\begin{center}
			\includegraphics[width=6.3cm,height=4.3cm]{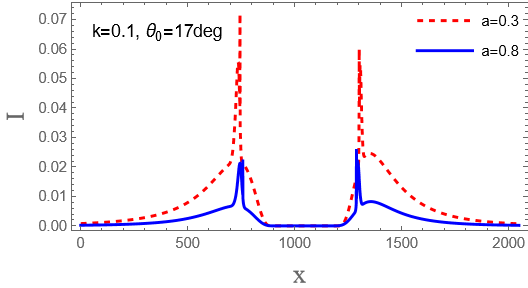}
			\includegraphics[width=6.7cm,height=4.3cm]{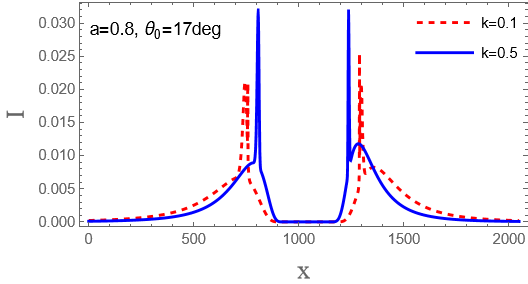}
			\parbox[c]{15.0cm}{\footnotesize{\bf Fig~12.}  
				The same Fig~10 but for for the BH with a retrograde accretion.}
			\label{fig12}
		\end{center}
		
		Figures 9 and 11 present synthetic 230 GHz images of the accretion disk around a BH embedded in PFDM, corresponding to prograde and retrograde configurations, respectively. For prograde scenarios, the observed radiation intensity increases significantly with higher spin parameter $a$, accompanied by a pronounced leftward concentration of high-intensity emission due to relativistic beaming effects. This asymmetry, resembling a characteristic D-shaped morphology, becomes more distinct as $a$ approaches extremal values. Simultaneously, increasing the PFDM parameter $k$ from 0.1 to 0.5 causes inward contraction of photon trajectories, consistent with the observed reduction in shadow diameter. Figure 10 quantifies the radiation intensity profiles along the $x$ and $y$ axes, confirming the spatial redistribution of emission under varying $a$ and $k$. In retrograde configurations, the radiation intensity decreases systematically with increasing $a$, while higher $k$ values induce a subtle but measurable enhancement of intensity. Figure 12 illustrates this $k$ dependent intensity shift along the $x$ axis, showing gradual displacement of the emission peak toward the BH center as $k$ increases. At 86 GHz (Figure 13), the lower angular resolution obscures fine-scale intensity variations, resulting in a saturated disk morphology that lacks the structural details resolvable at 230 GHz. This resolution-dependent contrast reduction underscores the importance of high-frequency observations for probing horizon-scale accretion dynamics.
		\begin{center}
			\begin{minipage}[b]{0.79\textwidth}
				\includegraphics[width=\textwidth,height=3.7cm]{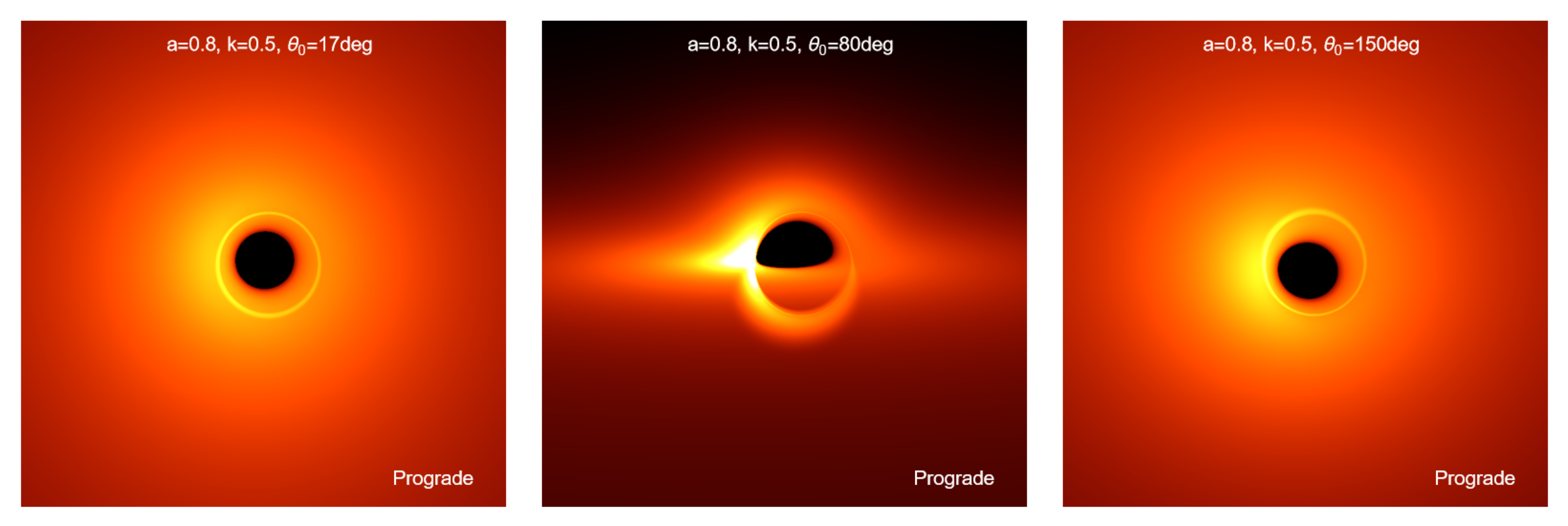}\\
				\vspace{0.5cm}
				\includegraphics[width=\textwidth,height=3.7cm]{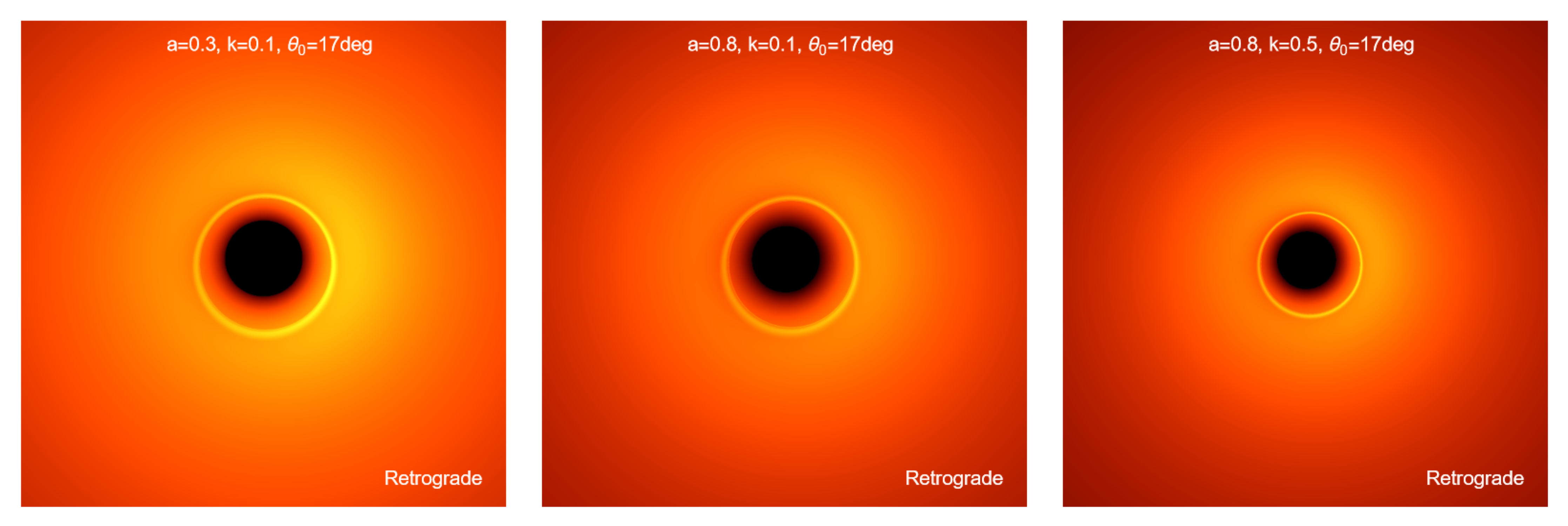}
			\end{minipage}
			\begin{minipage}[b]{0.2\textwidth}
				\includegraphics[width=0.4\textwidth,height=7.2cm]{fbar.png}
			\end{minipage}
			\parbox[c]{15.0cm}{\footnotesize{\bf Fig~13.}  
				Synthetic images of the BH accretion disk at 86 GHz ($M=1$). Top Panel: Prograde images for fixed parameters $(a, k) = (0.8, 0.5)$ at various inclination angles. Bottom Panel: Retrograde images at a fixed inclination angle of $17^{\circ}$ for different parameter sets $(a, k)$.}
			\label{fig13}
		\end{center}
		
		\subsection{Confronting the Model with EHT Observations: Constraints and Comparison with M87$^{*}$}
		\label{sec:3-2}
		
		\par
		To place quantitative constraints on the PFDM model using astrophysical BH observations, we apply our theoretical framework to both M87$^{*}$ and Sgr A$^{*}$, deriving constraints on the PFDM parameter $k$ from their measured shadow diameters. The Event Horizon Telescope collaboration has provided precise measurements of the shadow diameters for both BHs. For M87$^{*}$, the angular shadow diameter is $\delta=(42 {\pm} 3)$ ${\rm {\mu as}}$ with a distance of $D = 16.8$ Mpc and BH mass $M_{BH} = 6.5 \times 10^9 M_\odot$ \cite{EventHorizonTelescope:2019dse,akiyama2019first}. For Sgr A$^{*}$, the measured values are $\delta = (48.7 \pm 7)$${\rm {\mu as}}$, $D = 8.35$ kpc, and $M_{BH} = 4.3 \times 10^6 M_\odot$ \cite{EventHorizonTelescope:2022wkp}. In our theoretical approach, we employ a geometrically robust characterization of BH shadows that properly accounts for frame-dragging induced deformations in rotating spacetimes. Following the established methodology in \cite{kumar2020black}, we compute the shadow area $\mathcal{D}$ through parametric integration of the photon region boundary:
		\begin{equation}
			\label{eq:41}
			\mathcal{D}=2 \int \beta(r_{p}){\rm d}\alpha(r_{p})=2 \int^{r_{p}^{+}}_{r_{p}^{-}}(\beta(r_{p}){\rm d}\frac{\alpha(r_{p})}{{\rm d}r_{p}}){\rm d}r_{p}.
		\end{equation}
		where $r_p^-$ and $r_p^+$ denote the minimum and maximum radii of the photon region, and $(\alpha, \beta)$ represent the celestial coordinates on the observer's sky. The shadow's equivalent diameter is then defined as $d_{\rm DM} = 2\sqrt{\mathcal{D}/\pi}$, corresponding to the diameter of a circle with the same area as the shadow  \cite{abdujabbarov2015coordinate}. This area-based measure provides a coordinate-independent and shape-agnostic characterization of the shadow size, as it does not rely on any prior assumptions about the shadow's shape or the location of its center. The conversion to angular diameter follows the standard relativistic prescription: $\theta_{\rm DM} = d_{\rm DM} / D \times (180/\pi) \times 3600 \times 10^6~\mu{\rm as}$. 
		\begin{center}
			\includegraphics[width=6.8cm,height=5cm]{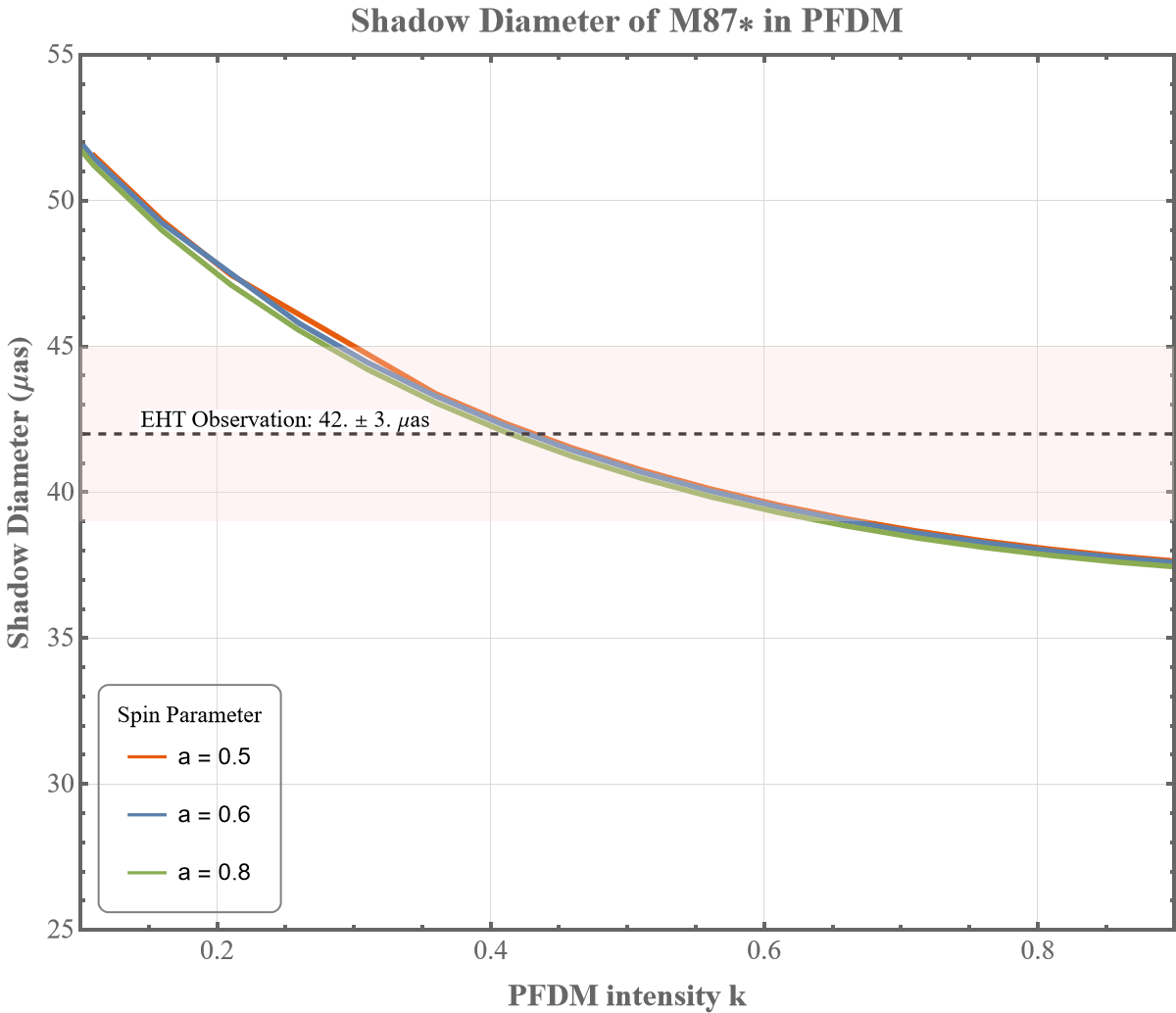}
			\includegraphics[width=6.8cm,height=5cm]{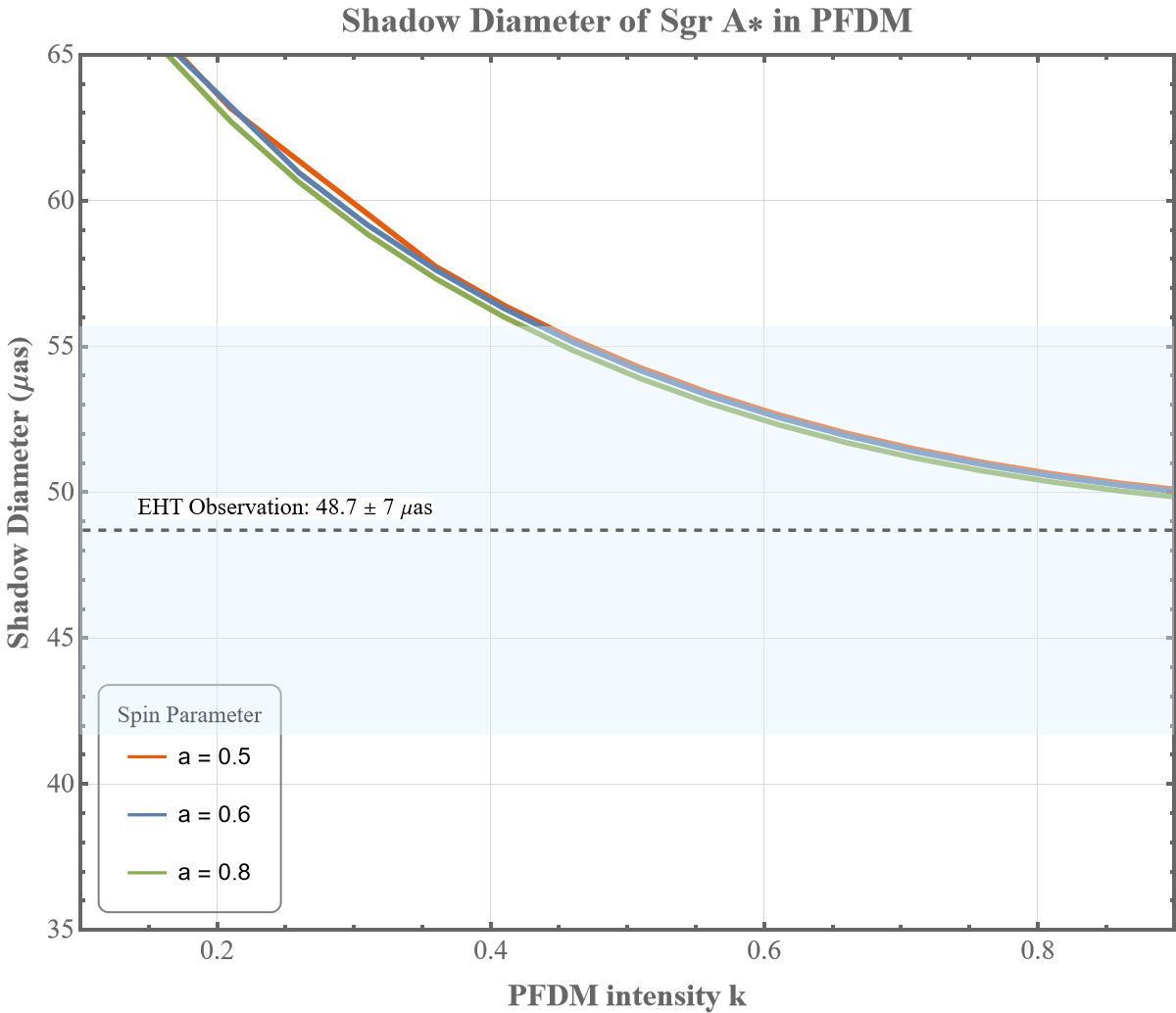}
			\parbox[c]{15.0cm}{\footnotesize{\bf Fig~14.}  
				Comparison of theoretical predictions with EHT shadow diameter measurements for (Left) M87$^{*}$ and (Right) Sgr A$^{*}$. The horizontal bands indicate the observational ranges of $42\pm3~\mu{\rm as}$ for M87* and $48.7\pm7~\mu{\rm as}$ for Sgr A$^{*}$. The curves show the theoretical dependence on the PFDM parameter $k$ for spin values $a=0.5,0.6,0.8$.}
			\label{fig14}
		\end{center}
		
		Figure 14 displays the variation of theoretical shadow diameter with the PFDM parameter $k$ for spin values $a = 0.5, 0.6, 0.8$, compared against EHT observational constraints. For M87$^{*}$, the values of $k$ where the theoretical predictions enter the EHT observational range are $k = 0.297$ for $a=0.5$, $k = 0.289$ for $a=0.6$, and $k = 0.278$ for $a=0.8$. For Sgr A$^{*}$, the corresponding values are $k = 0.441$, $k = 0.440$, and $k = 0.419$ for the same spin parameters. Considering the previously established theoretical constraint $k < 0.5$ from viability conditions of the PFDM metric, our analysis restricts $k$ to the ranges $0.278 \leq k \leq 0.5$ for M87$^{*}$ and $0.419 \leq k \leq 0.5$ for Sgr A$^{*}$. Owing to the narrower allowed interval for Sgr~A$^{*}$ and its stronger sensitivity to observational uncertainties, we focus our subsequent discussion primarily on M87$^{*}$, which provides more robust constraints on the PFDM model. Our systematic parameter space analysis reveals a remarkably weak sensitivity of the shadow diameter to the spin parameter $a$. This insensitivity validates our approach of presenting comprehensive results across multiple spin values.

		\par
		To more intuitively illustrate how the presence of dark matter affects the appearance of the accretion disk, and to explore whether the M87$^{*}$ BH could plausibly be described by such a scenario, we blurred our simulated images at 230 GHz using a Gaussian filter with a standard deviation corresponding to a field of view of 1/12, thereby approximating the nominal resolution of the EHT \cite{Gralla:2019xty}. Figure 15 shows the observed EHT image of M87$^{*}$, our processed Kerr BH image, and BH images in PFDM for different values of the dark matter parameter $k$ with fixed spin parameter $a=0.8$ and inclination $\theta_{0}=17^{\circ}$.
		\begin{center}
			\includegraphics[width=6.5cm,height=6.5cm]{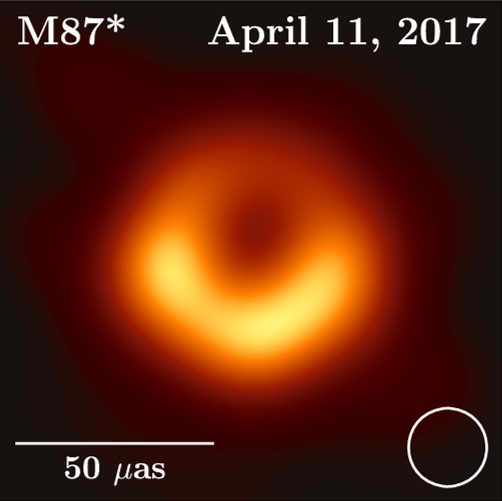}
			\includegraphics[width=6.5cm,height=6.5cm]{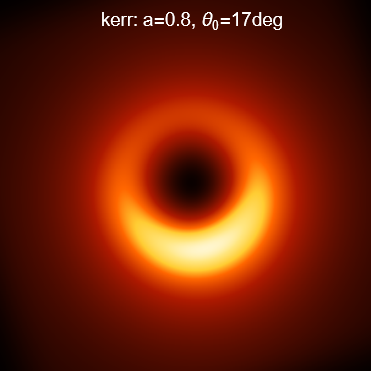}
			\includegraphics[width=6.5cm,height=6.5cm]{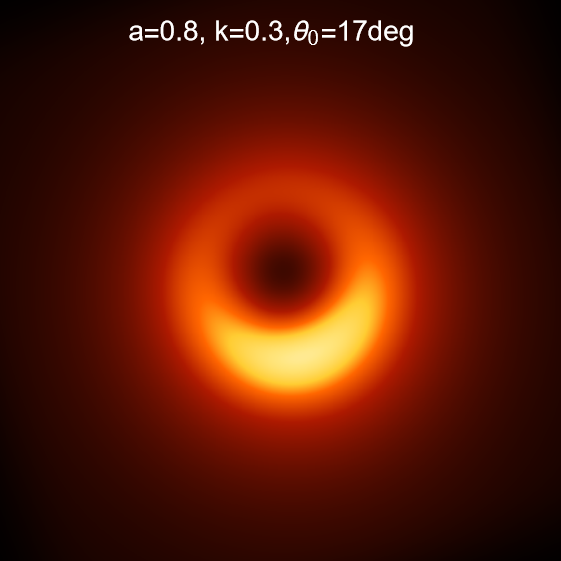}
			\includegraphics[width=6.5cm,height=6.5cm]{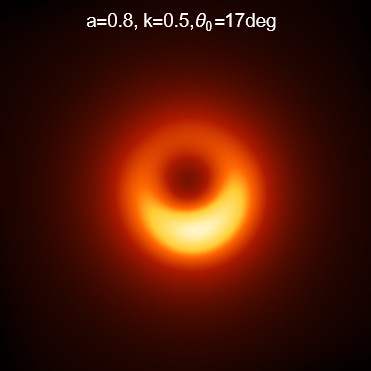}
			\parbox[c]{15.0cm}{\footnotesize{\bf Fig~15.}  
				Simulated 230 GHz images blurred to match the nominal EHT resolution, comparing different dark matter scenarios with the observed M87$^{*}$ image (top left; adapted from \cite{EventHorizonTelescope:2019dse}). The top right panel shows the processed image of a Kerr BH with spin parameter $a=0.8$ and inclination angle $\theta_{0}=17^{\circ}$. The bottom panels correspond to BHs embedded in PFDM with the same spin and inclination but increasing PFDM intensity: $k=0.3$ (bottom left) and $k=0.5$ (bottom right).}
			\label{fig15}
		\end{center}
		\par
		It can be seen that when dark matter is included, the shadow region shrinks significantly as the PFDM strength $k$ increases, while the area of bright emission in the surrounding accretion disk expands accordingly. This behavior demonstrates that the presence of PFDM can effectively alter both the apparent size of the BH shadow and the brightness structure of the disk, offering additional degrees of freedom that can be tuned to match observed features. Although our simplified model does not capture all fine structures-such as the asymmetric brightness and turbulence seen in the EHT data-these results demonstrate that PFDM can naturally reproduce the observed compact shadow size and crescent-like morphology under certain parameter ranges.
		
		\par
		Therefore, BHs embedded in PFDM provide a physically motivated alternative to the standard Kerr scenario for M87$^{*}$, highlighting the importance of including dark matter effects in realistic GRMHD modeling and parameter inference.

		\section{Conclusions and Outlook}
		\label{sec:4}
		We present a systematic investigation of the spacetime properties of rotating BHs immersed in PFDM. Particular emphasis is placed on examining the joint influence of the BH spin parameter $a$ and PFDM intensity parameter $k$ on particle dynamics, photon trajectories, and shadow morphology, with our analysis informed by the morphological insights from the M87$^{*}$ observation.

		Initially, we constrain the physically allowed ranges of $a$ and $k$ by enforcing the existence of an event horizon through metric conditions.  By solving the photon sphere equations, we determine the BH shadow structure and demonstrate pronounced modifications induced by the presence of dark matter.  Specifically, increases in the parameter $k$ yield measurable reductions in shadow size and induce notable shape variations, while the BH spin parameter $a$ further modulates these features by shifting both the event horizon and ISCO.

		Utilizing a fisheye camera ray-tracing methodology, we produce high-resolution synthetic images of BHs embedded in PFDM, quantitatively assessing angular-dependent photon trajectories and associated Doppler shift distributions. To ensure robust and comprehensive results, we investigate both prograde and retrograde accretion scenarios and consider multiple observational frequencies at 230 GHz and 86 GHz to evaluate resolution-dependent observational features. Our analyses illustrate that the inclusion of PFDM systematically reshapes the apparent shadow size of the BH and enhances the bright emission region of the accretion disk, introducing an additional physically motivated parameter that connects theoretical modeling with high-resolution VLBI observations.

		By combining constraints from shadow morphology calculations with angular diameter measurements of M87$^{*}$ and Sgr A$^{*}$ from the Event Horizon Telescope, we derive observationally consistent bounds on the PFDM parameter space. Direct comparison with EHT images of M87* demonstrates that incorporating dark matter provides a natural explanation for both the compact shadow and the extended crescent-shaped emission observed. Although our model relies on a geometrically thin disk approximation and does not include detailed radiative transfer, our results provide compelling evidence that dark matter can play a non-negligible role in shaping horizon-scale observables around supermassive BHs.

		Collectively, our results highlight that BHs immersed in PFDM environments provide a physically plausible and observationally consistent alternative for modeling M87$^{*}$ and analogous astrophysical systems. This underscores the critical importance of incorporating dark matter effects into future General GRMHD simulations and event horizon imaging analyses. An especially intriguing aspect of PFDM spacetimes is the possible emergence of multiple ISCO solutions within the same radial domain, which suggests a richer class of accretion configurations and dynamical transitions beyond the standard single-ISCO paradigm. Exploring the stability, accretion dynamics, and observational imprints associated with such multi-ISCO structures constitutes a natural and promising direction for our future work. Future extensions of this framework will include realistic accretion disk geometries, self-consistent radiative transfer modeling, and dynamic dark matter interactions, thereby refining constraints on dark matter properties through precision observations of BH shadows.
		
		\section*{Acknowledgments}
		This work is supported by the National Key R\&D Program (No. 2024YFA161100), and National Natural Science Foundation of China (grant Nos. 12133003). E. W. L. is also supported by the Guangxi Talent Program (``Highland of Innovation Talents''). K. L. was supported by Fapesq-PB of Brazil.
		\bibliographystyle{elsarticle-num} 
		\bibliography{dm}

@article{Bahcall:1991qs,
    author = "Bahcall, John N. and Flynn, Chris and Gould, Andrew",
    title = "{Local dark matter from a carefully selected sample}",
    reportNumber = "IASSNS-AST-91-7",
    doi = "10.1086/171201",
    journal = "Astrophys. J.",
    volume = "389",
    pages = "234--250",
    year = "1992"
}

@article{Zaritsky:1996ch,
    author = "Zaritsky, Dennis and Smith, Rodney and Frenk, Carlos and White, Simon D. M.",
    title = "{More satellites of spiral galaxies}",
    eprint = "astro-ph/9611199",
    archivePrefix = "arXiv",
    doi = "10.1086/303784",
    journal = "Astrophys. J.",
    volume = "478",
    pages = "39--48",
    year = "1997"
}

@article{Mateo:1998wg,
    author = "Mateo, Mario",
    title = "{Dwarf galaxies of the Local Group}",
    eprint = "astro-ph/9810070",
    archivePrefix = "arXiv",
    doi = "10.1146/annurev.astro.36.1.435",
    journal = "Ann. Rev. Astron. Astrophys.",
    volume = "36",
    pages = "435--506",
    year = "1998"
}

@article{Koopmans:2002qh,
    author = "Koopmans, L. V. E. and Treu, T.",
    title = "{The structure and dynamics of luminous and dark matter in the early-type lens galaxy of 0047-281 at z=0.485}",
    eprint = "astro-ph/0205281",
    archivePrefix = "arXiv",
    doi = "10.1086/345423",
    journal = "Astrophys. J.",
    volume = "583",
    pages = "606--615",
    year = "2003"
}

@article{kiselev2003quintessential,
  title={Quintessential solution of dark matter rotation curves and its simulation by extra dimensions},
  author={Kiselev, VV},
  journal={arXiv preprint gr-qc/0303031},
  year={2003}
}

@article{Kiselev:2002dx,
    author = "Kiselev, V. V.",
    title = "{Quintessence and black holes}",
    eprint = "gr-qc/0210040",
    archivePrefix = "arXiv",
    doi = "10.1088/0264-9381/20/6/310",
    journal = "Class. Quant. Grav.",
    volume = "20",
    pages = "1187--1198",
    year = "2003"
}

@article{Li:2012zx,
    author = "Li, Ming-Hsun and Yang, Kwei-Chou",
    title = "{Galactic Dark Matter in the Phantom Field}",
    eprint = "1204.3178",
    archivePrefix = "arXiv",
    primaryClass = "astro-ph.CO",
    reportNumber = "CYCU-HEP-12-04",
    doi = "10.1103/PhysRevD.86.123015",
    journal = "Phys. Rev. D",
    volume = "86",
    pages = "123015",
    year = "2012"
}

@article{Xu:2017bpz,
    author = "Xu, Zhaoyi and Wang, Jiancheng and Hou, Xian",
    title = "{Kerr\textendash{}anti-de Sitter/de Sitter black hole in perfect fluid dark matter background}",
    eprint = "1711.04538",
    archivePrefix = "arXiv",
    primaryClass = "gr-qc",
    doi = "10.1088/1361-6382/aabcb6",
    journal = "Class. Quant. Grav.",
    volume = "35",
    number = "11",
    pages = "115003",
    year = "2018"
}

@article{Rizwan:2018rgs,
    author = "Rizwan, Muhammad and Jamil, Mubasher and Jusufi, Kimet",
    title = "{Distinguishing a Kerr-like black hole and a naked singularity in perfect fluid dark matter via precession frequencies}",
    eprint = "1812.01331",
    archivePrefix = "arXiv",
    primaryClass = "gr-qc",
    doi = "10.1103/PhysRevD.99.024050",
    journal = "Phys. Rev. D",
    volume = "99",
    number = "2",
    pages = "024050",
    year = "2019"
}

@article{Narzilloev:2020qtd,
    author = "Narzilloev, Bakhtiyor and Rayimbaev, Javlon and Shaymatov, Sanjar and Abdujabbarov, Ahmadjon and Ahmedov, Bobomurat and Bambi, Cosimo",
    title = "{Dynamics of test particles around a Bardeen black hole surrounded by perfect fluid dark matter}",
    eprint = "2011.06148",
    archivePrefix = "arXiv",
    primaryClass = "gr-qc",
    doi = "10.1103/PhysRevD.102.104062",
    journal = "Phys. Rev. D",
    volume = "102",
    number = "10",
    pages = "104062",
    year = "2020"
}

@article{Cao:2021dcq,
    author = "Cao, Yihe and Feng, Hanwen and Hong, Wei and Tao, Jun",
    title = "{Joule\textendash{}Thomson expansion of RN-AdS black hole immersed in perfect fluid dark matter}",
    eprint = "2101.08199",
    archivePrefix = "arXiv",
    primaryClass = "gr-qc",
    doi = "10.1088/1572-9494/ac1066",
    journal = "Commun. Theor. Phys.",
    volume = "73",
    number = "9",
    pages = "095403",
    year = "2021"
}

@article{Zhou:2022eft,
    author = "Zhou, Xingyu and Xue, Yadong and Mu, Benrong and Tao, Jun",
    title = "{Temporal and spatial chaos of RN-AdS black holes immersed in Perfect Fluid Dark Matter}",
    eprint = "2209.03612",
    archivePrefix = "arXiv",
    primaryClass = "gr-qc",
    reportNumber = "CTP-SCU/2022010",
    doi = "10.1016/j.dark.2023.101168",
    journal = "Phys. Dark Univ.",
    volume = "39",
    pages = "101168",
    year = "2023"
}

@article{Rizwan:2023ivp,
    author = "Rizwan, Muhammad and Jusufi, Kimet",
    title = "{Topological classes of thermodynamics of black holes in perfect fluid dark matter background}",
    eprint = "2310.15182",
    archivePrefix = "arXiv",
    primaryClass = "gr-qc",
    doi = "10.1140/epjc/s10052-023-12126-1",
    journal = "Eur. Phys. J. C",
    volume = "83",
    number = "10",
    pages = "944",
    year = "2023"
}

@article{EventHorizonTelescope:2019dse,
    author = "Akiyama, Kazunori and others",
    collaboration = "Event Horizon Telescope",
    title = "{First M87 Event Horizon Telescope Results. I. The Shadow of the Supermassive Black Hole}",
    eprint = "1906.11238",
    archivePrefix = "arXiv",
    primaryClass = "astro-ph.GA",
    doi = "10.3847/2041-8213/ab0ec7",
    journal = "Astrophys. J. Lett.",
    volume = "875",
    pages = "L1",
    year = "2019"
}

@article{EventHorizonTelescope:2022wkp,
    author = "Akiyama, Kazunori and others",
    collaboration = "Event Horizon Telescope",
    title = "{First Sagittarius A* Event Horizon Telescope Results. I. The Shadow of the Supermassive Black Hole in the Center of the Milky Way}",
    eprint = "2311.08680",
    archivePrefix = "arXiv",
    primaryClass = "astro-ph.HE",
    doi = "10.3847/2041-8213/ac6674",
    journal = "Astrophys. J. Lett.",
    volume = "930",
    number = "2",
    pages = "L12",
    year = "2022"
}

@article{Lu:2023bbn,
    author = "Lu, Ru-Sen and others",
    title = "{A ring-like accretion structure in M87 connecting its black hole and jet}",
    eprint = "2304.13252",
    archivePrefix = "arXiv",
    primaryClass = "astro-ph.HE",
    doi = "10.1038/s41586-023-05843-w",
    journal = "Nature",
    volume = "616",
    number = "7958",
    pages = "686--690",
    year = "2023"
}

@article{janssen2021event,
  title={Event Horizon Telescope observations of the jet launching and collimation in Centaurus A},
  author={Janssen, Michael and Falcke, Heino and Kadler, Matthias and Ros, Eduardo and Wielgus, Maciek and Akiyama, Kazunori and Balokovi{\'c}, Mislav and Blackburn, Lindy and Bouman, Katherine L and Chael, Andrew and others},
  journal={Nature Astronomy},
  volume={5},
  number={10},
  pages={1017--1028},
  year={2021},
  publisher={Nature Publishing Group UK London}
}

@article{akiyama2019first,
  title={First M87 event horizon telescope results. IV. Imaging the central supermassive black hole},
  author={Akiyama, Kazunori and Alberdi, Antxon and Alef, Walter and Asada, Keiichi and Azulay, Rebecca and Baczko, Anne-Kathrin and Ball, David and Balokovi{\'c}, Mislav and Barrett, John and Bintley, Dan and others},
  journal={The Astrophysical Journal Letters},
  volume={875},
  number={1},
  pages={L4},
  year={2019},
  publisher={IOP Publishing}
}

@article{EventHorizonTelescope:2022wok,
    author = "Akiyama, Kazunori and others",
    collaboration = "Event Horizon Telescope",
    title = "{First Sagittarius A* Event Horizon Telescope Results. III. Imaging of the Galactic Center Supermassive Black Hole}",
    eprint = "2311.09479",
    archivePrefix = "arXiv",
    primaryClass = "astro-ph.HE",
    doi = "10.3847/2041-8213/ac6429",
    journal = "Astrophys. J. Lett.",
    volume = "930",
    number = "2",
    pages = "L14",
    year = "2022"
}

@article{EventHorizonTelescope:2022urf,
    author = "Akiyama, Kazunori and others",
    collaboration = "Event Horizon Telescope",
    title = "{First Sagittarius A* Event Horizon Telescope Results. V. Testing Astrophysical Models of the Galactic Center Black Hole}",
    eprint = "2311.09478",
    archivePrefix = "arXiv",
    primaryClass = "astro-ph.HE",
    reportNumber = "FERMILAB-PUB-22-419-PPD",
    doi = "10.3847/2041-8213/ac6672",
    journal = "Astrophys. J. Lett.",
    volume = "930",
    number = "2",
    pages = "L16",
    year = "2022"
}

@article{ripperda2022black,
  title={Black hole flares: ejection of accreted magnetic flux through 3D plasmoid-mediated reconnection},
  author={Ripperda, Bart and Liska, Matthew and Chatterjee, Koushik and Musoke, Gibwa and Philippov, Alexander A and Markoff, Sera B and Tchekhovskoy, Alexander and Younsi, Ziri},
  journal={The Astrophysical Journal Letters},
  volume={924},
  number={2},
  pages={L32},
  year={2022},
  publisher={IOP Publishing}
}

@article{Yuan:2022mkw,
    author = "Yuan, Feng and Wang, Haiyang and Yang, Hai",
    title = "{The Accretion Flow in M87 is Really MAD}",
    eprint = "2201.00512",
    archivePrefix = "arXiv",
    primaryClass = "astro-ph.HE",
    doi = "10.3847/1538-4357/ac4714",
    journal = "Astrophys. J.",
    volume = "924",
    number = "2",
    pages = "124",
    year = "2022"
}

@article{nalewajko2024chaotic,
  title={Chaotic magnetic disconnections trigger flux eruptions in accretion flows channeled onto magnetically saturated Kerr black holes},
  author={Nalewajko, Krzysztof and Kapusta, Mateusz and Janiuk, Agnieszka},
  journal={Astronomy \& Astrophysics},
  volume={692},
  pages={A37},
  year={2024},
  publisher={EDP Sciences}
}

@article{yang2024modeling,
  title={Modeling the inner part of the jet in M87: Confronting jet morphology with theory},
  author={Yang, Hai and Yuan, Feng and Li, Hui and Mizuno, Yosuke and Guo, Fan and Lu, Rusen and Ho, Luis C and Lin, Xi and Zdziarski, Andrzej A and Wang, Jieshuang},
  journal={Science Advances},
  volume={10},
  number={12},
  pages={eadn3544},
  year={2024},
  publisher={American Association for the Advancement of Science}
}

@article{akiyama2022first,
  title={First Sagittarius A* event horizon telescope results. VI. Testing the black hole metric},
  author={Akiyama, Kazunori and Alberdi, Antxon and Alef, Walter and Algaba, Juan Carlos and Anantua, Richard and Asada, Keiichi and Azulay, Rebecca and Bach, Uwe and Baczko, Anne-Kathrin and Ball, David and others},
  journal={The Astrophysical Journal Letters},
  volume={930},
  number={2},
  pages={L17},
  year={2022},
  publisher={IOP Publishing}
}

@article{cunha2020lensing,
  title={Lensing and shadow of a black hole surrounded by a heavy accretion disk},
  author={Cunha, Pedro VP and Eir{\'o}, Nelson A and Herdeiro, Carlos AR and Lemos, Jos{\'e} PS},
  journal={Journal of Cosmology and Astroparticle Physics},
  volume={2020},
  number={03},
  pages={035},
  year={2020},
  publisher={IOP Publishing}
}

@article{Zeng:2020dco,
    author = "Zeng, Xiao-Xiong and Zhang, Hai-Qing and Zhang, Hongbao",
    title = "{Shadows and photon spheres with spherical accretions in the four-dimensional Gauss\textendash{}Bonnet black hole}",
    eprint = "2004.12074",
    archivePrefix = "arXiv",
    primaryClass = "gr-qc",
    doi = "10.1140/epjc/s10052-020-08449-y",
    journal = "Eur. Phys. J. C",
    volume = "80",
    number = "9",
    pages = "872",
    year = "2020"
}

@article{Ma:2019ybz,
    author = "Ma, Liang and Lu, H.",
    title = "{Bounds on photon spheres and shadows of charged black holes in Einstein-Gauss-Bonnet-Maxwell gravity}",
    eprint = "1912.05569",
    archivePrefix = "arXiv",
    primaryClass = "gr-qc",
    doi = "10.1016/j.physletb.2020.135535",
    journal = "Phys. Lett. B",
    volume = "807",
    pages = "135535",
    year = "2020"
}

@article{Chen:2020qyp,
    author = "Chen, Songbai and Wang, Mingzhi and Jing, Jiliang",
    title = "{Polarization effects in Kerr black hole shadow due to the coupling between photon and bumblebee field}",
    eprint = "2004.08857",
    archivePrefix = "arXiv",
    primaryClass = "gr-qc",
    doi = "10.1007/JHEP07(2020)054",
    journal = "JHEP",
    volume = "07",
    pages = "054",
    year = "2020"
}

@article{Zhang:2021hit,
    author = "Zhang, Zelin and Chen, Songbai and Qin, Xin and Jing, Jiliang",
    title = "{Polarized image of a Schwarzschild black hole with a thin accretion disk as photon couples to Weyl tensor}",
    eprint = "2106.07981",
    archivePrefix = "arXiv",
    primaryClass = "gr-qc",
    doi = "10.1140/epjc/s10052-021-09786-2",
    journal = "Eur. Phys. J. C",
    volume = "81",
    number = "11",
    pages = "991",
    year = "2021"
}

@article{Gan:2021pwu,
    author = "Gan, Qingyu and Wang, Peng and Wu, Houwen and Yang, Haitang",
    title = "{Photon spheres and spherical accretion image of a hairy black hole}",
    eprint = "2104.08703",
    archivePrefix = "arXiv",
    primaryClass = "gr-qc",
    doi = "10.1103/PhysRevD.104.024003",
    journal = "Phys. Rev. D",
    volume = "104",
    number = "2",
    pages = "024003",
    year = "2021"
}

@article{Peng:2020wun,
    author = "Peng, Jun and Guo, Minyong and Feng, Xing-Hui",
    title = "{Influence of quantum correction on black hole shadows, photon rings, and lensing rings}",
    eprint = "2008.00657",
    archivePrefix = "arXiv",
    primaryClass = "gr-qc",
    doi = "10.1088/1674-1137/ac06bb",
    journal = "Chin. Phys. C",
    volume = "45",
    number = "8",
    pages = "085103",
    year = "2021"
}

@article{Zeng:2021mok,
    author = "Zeng, Xiao-Xiong and He, Ke-Jian and Li, Guo-Ping",
    title = "{Effects of dark matter on shadows and rings of Brane-World black holes illuminated by various accretions}",
    eprint = "2111.05090",
    archivePrefix = "arXiv",
    primaryClass = "gr-qc",
    doi = "10.1007/s11433-022-1896-0",
    journal = "Sci. China Phys. Mech. Astron.",
    volume = "65",
    number = "9",
    pages = "290411",
    year = "2022"
}

@article{Guo:2021bhr,
    author = "Guo, Sen and Li, Guan-Ru and Liang, En-Wei",
    title = "{Influence of accretion flow and magnetic charge on the observed shadows and rings of the Hayward black hole}",
    eprint = "2112.11227",
    archivePrefix = "arXiv",
    primaryClass = "astro-ph.HE",
    doi = "10.1103/PhysRevD.105.023024",
    journal = "Phys. Rev. D",
    volume = "105",
    number = "2",
    pages = "023024",
    year = "2022"
}

@article{guo2022quasinormal,
  title={Quasinormal modes of black holes with multiple photon spheres},
  author={Guo, Guangzhou and Wang, Peng and Wu, Houwen and Yang, Haitang},
  journal={Journal of High Energy Physics},
  volume={2022},
  number={6},
  pages={1--24},
  year={2022},
  publisher={Springer}
}

@article{Kuang:2022ojj,
    author = "Kuang, Xiao-Mei and Tang, Zi-Yu and Wang, Bin and Wang, Anzhong",
    title = "{Constraining a modified gravity theory in strong gravitational lensing and black hole shadow observations}",
    eprint = "2206.05878",
    archivePrefix = "arXiv",
    primaryClass = "gr-qc",
    doi = "10.1103/PhysRevD.106.064012",
    journal = "Phys. Rev. D",
    volume = "106",
    number = "6",
    pages = "064012",
    year = "2022"
}

@article{Hou:2022eev,
    author = "Hou, Yehui and Zhang, Zhenyu and Yan, Haopeng and Guo, Minyong and Chen, Bin",
    title = "{Image of a Kerr-Melvin black hole with a thin accretion disk}",
    eprint = "2206.13744",
    archivePrefix = "arXiv",
    primaryClass = "gr-qc",
    doi = "10.1103/PhysRevD.106.064058",
    journal = "Phys. Rev. D",
    volume = "106",
    number = "6",
    pages = "064058",
    year = "2022"
}

@article{Huang:2023ilm,
    author = "Huang, Yu-Xiang and Guo, Sen and Cui, Yu-Hao and Jiang, Qing-Quan and Lin, Kai",
    title = "{Influence of accretion disk on the optical appearance of the Kazakov-Solodukhin black hole}",
    eprint = "2311.00302",
    archivePrefix = "arXiv",
    primaryClass = "gr-qc",
    doi = "10.1103/PhysRevD.107.123009",
    journal = "Phys. Rev. D",
    volume = "107",
    number = "12",
    pages = "123009",
    year = "2023"
}

@article{Meng:2024puu,
    author = "Meng, Yuan and Kuang, Xiao-Mei and Wang, Xi-Jing and Wang, Bin and Wu, Jian-Pin",
    title = {{Images of hairy Reissner\textendash{}Nordstr\"om black hole illuminated by static accretions}},
    eprint = "2401.05634",
    archivePrefix = "arXiv",
    primaryClass = "gr-qc",
    doi = "10.1140/epjc/s10052-024-12686-w",
    journal = "Eur. Phys. J. C",
    volume = "84",
    number = "3",
    pages = "305",
    year = "2024"
}

@article{Cui:2024wvz,
    author = "Cui, Yu-Hao and Guo, Sen and Huang, Yu-Xiang and Liang, Yu and Lin, Kai",
    title = "{Optical appearance of numerical black hole solutions in higher derivative gravity}",
    eprint = "2408.03387",
    archivePrefix = "arXiv",
    primaryClass = "gr-qc",
    doi = "10.1140/epjc/s10052-024-13153-2",
    journal = "Eur. Phys. J. C",
    volume = "84",
    number = "8",
    pages = "772",
    year = "2024"
}

@article{hu2024influences,
  title={Influences of tilted thin accretion disks on the observational appearance of hairy black holes in Horndeski gravity},
  author={Hu, Shiyang and Li, Dan and Deng, Chen and Wu, Xin and Liang, Enwei},
  journal={Journal of Cosmology and Astroparticle Physics},
  volume={2024},
  number={04},
  pages={089},
  year={2024},
  publisher={IOP Publishing}
}

@article{zhang2024imaging,
  title={Imaging thick accretion disks and jets surrounding black holes},
  author={Zhang, Zhenyu and Hou, Yehui and Guo, Minyong and Chen, Bin},
  journal={Journal of Cosmology and Astroparticle Physics},
  volume={2024},
  number={05},
  pages={032},
  year={2024},
  publisher={IOP Publishing}
}

@article{Haroon:2018ryd,
    author = "Haroon, Sumarna and Jamil, Mubasher and Jusufi, Kimet and Lin, Kai and Mann, Robert B.",
    title = "{Shadow and Deflection Angle of Rotating Black Holes in Perfect Fluid Dark Matter with a Cosmological Constant}",
    eprint = "1810.04103",
    archivePrefix = "arXiv",
    primaryClass = "gr-qc",
    doi = "10.1103/PhysRevD.99.044015",
    journal = "Phys. Rev. D",
    volume = "99",
    number = "4",
    pages = "044015",
    year = "2019"
}

@article{Heydari-Fard:2022xhr,
    author = "Heydari-Fard, Malihe and Honarvar, Sara Ghassemi and Heydari-Fard, Mohaddese",
    title = "{Thin accretion disc luminosity and its image around rotating black holes in perfect fluid dark matter}",
    eprint = "2210.04173",
    archivePrefix = "arXiv",
    primaryClass = "gr-qc",
    doi = "10.1093/mnras/stad558",
    journal = "Mon. Not. Roy. Astron. Soc.",
    volume = "521",
    number = "1",
    pages = "708--716",
    year = "2023"
}

@article{EventHorizonTelescope:2021btj,
    author = "Akiyama, Kazunori and others",
    collaboration = "Event Horizon Telescope",
    title = "{The Polarized Image of a Synchrotron-emitting Ring of Gas Orbiting a Black Hole}",
    eprint = "2105.01804",
    archivePrefix = "arXiv",
    primaryClass = "astro-ph.HE",
    reportNumber = "FERMILAB-PUB-21-848-PPD",
    doi = "10.3847/1538-4357/abf117",
    journal = "Astrophys. J.",
    volume = "912",
    number = "1",
    pages = "35",
    year = "2021"
}

@article{kumar2020black,
  title={Black hole parameter estimation from its shadow},
  author={Kumar, Rahul and Ghosh, Sushant G},
  journal={The Astrophysical Journal},
  volume={892},
  number={2},
  pages={78},
  year={2020},
  publisher={IOP Publishing}
}

@article{abdujabbarov2015coordinate,
  title={A coordinate-independent characterization of a black hole shadow},
  author={Abdujabbarov, AA and Rezzolla, L and Ahmedov, BJ},
  journal={Monthly Notices of the Royal Astronomical Society},
  volume={454},
  number={3},
  pages={2423--2435},
  year={2015},
  publisher={Oxford University Press}
}

@article{Hu:2020usx,
    author = "Hu, Zezhou and Zhong, Zhen and Li, Peng-Cheng and Guo, Minyong and Chen, Bin",
    title = "{QED effect on a black hole shadow}",
    eprint = "2012.07022",
    archivePrefix = "arXiv",
    primaryClass = "gr-qc",
    doi = "10.1103/PhysRevD.103.044057",
    journal = "Phys. Rev. D",
    volume = "103",
    number = "4",
    pages = "044057",
    year = "2021"
}

@article{Gralla:2019drh,
    author = "Gralla, Samuel E. and Lupsasca, Alexandru",
    title = "{Lensing by Kerr Black Holes}",
    eprint = "1910.12873",
    archivePrefix = "arXiv",
    primaryClass = "gr-qc",
    doi = "10.1103/PhysRevD.101.044031",
    journal = "Phys. Rev. D",
    volume = "101",
    number = "4",
    pages = "044031",
    year = "2020"
}

@article{Lindquist:1966igj,
    author = "Lindquist, Richard W.",
    title = "{Relativistic transport theory}",
    doi = "10.1016/0003-4916(66)90207-7",
    journal = "Annals Phys.",
    volume = "37",
    number = "3",
    pages = "487--518",
    year = "1966"
}

@article{cunningham1975effects,
  title={The effects of redshifts and focusing on the spectrum of an accretion disk around a Kerr black hole},
  author={Cunningham, CT},
  journal={Astrophysical Journal, vol. 202, Dec. 15, 1975, pt. 1, p. 788-802.},
  volume={202},
  pages={788--802},
  year={1975}
}

@book{rybicki2024radiative,
  title={Radiative processes in astrophysics},
  author={Rybicki, George B and Lightman, Alan P},
  year={2024},
  publisher={John Wiley \& Sons}
}

@article{Chael:2021rjo,
    author = "Chael, Andrew and Johnson, Michael D. and Lupsasca, Alexandru",
    title = "{Observing the Inner Shadow of a Black Hole: A Direct View of the Event Horizon}",
    eprint = "2106.00683",
    archivePrefix = "arXiv",
    primaryClass = "astro-ph.HE",
    doi = "10.3847/1538-4357/ac09ee",
    journal = "Astrophys. J.",
    volume = "918",
    number = "1",
    pages = "6",
    year = "2021"
}

@article{EventHorizonTelescope:2019jan,
    author = "Akiyama, Kazunori and others",
    collaboration = "Event Horizon Telescope",
    title = "{First M87 Event Horizon Telescope Results. III. Data Processing and Calibration}",
    eprint = "1906.11240",
    archivePrefix = "arXiv",
    primaryClass = "astro-ph.GA",
    doi = "10.3847/2041-8213/ab0c57",
    journal = "Astrophys. J. Lett.",
    volume = "875",
    number = "1",
    pages = "L3",
    year = "2019"
}

@article{Gralla:2019xty,
    author = "Gralla, Samuel E. and Holz, Daniel E. and Wald, Robert M.",
    title = "{Black Hole Shadows, Photon Rings, and Lensing Rings}",
    eprint = "1906.00873",
    archivePrefix = "arXiv",
    primaryClass = "astro-ph.HE",
    doi = "10.1103/PhysRevD.100.024018",
    journal = "Phys. Rev. D",
    volume = "100",
    number = "2",
    pages = "024018",
    year = "2019"
}

@article{carter1968global,
  title={Global structure of the Kerr family of gravitational fields},
  author={Carter, Brandon},
  journal={Physical Review},
  volume={174},
  number={5},
  pages={1559},
  year={1968},
  publisher={APS}
}

@book{chandrasekhar1998mathematical,
  title={The mathematical theory of black holes},
  author={Chandrasekhar, Subrahmanyan},
  volume={69},
  year={1998},
  publisher={Oxford university press}
}

@article{wang2024image,
  title={Image of Kerr--de Sitter black holes illuminated by equatorial thin accretion disks},
  author={Wang, Ke and Feng, Chao-Jun and Wang, Towe},
  journal={The European Physical Journal C},
  volume={84},
  number={5},
  pages={457},
  year={2024},
  publisher={Springer}
}
		
		

	\end{document}